\documentclass[lettersize,journal]{IEEEtran}

\usepackage{amsmath,amsfonts}
\usepackage{algorithmic}
\usepackage{algorithm}
\usepackage{array}
\usepackage[caption=false,font=normalsize,labelfont=sf,textfont=sf]{subfig}
\usepackage{textcomp}
\usepackage{stfloats}
\usepackage{url}
\usepackage{verbatim}
\usepackage{graphicx}
\usepackage{cite}
\usepackage{balance}
\usepackage[table]{xcolor}
\usepackage{amsmath,amsfonts}

\usepackage{tabularx}
\usepackage{caption}
\usepackage{textcomp}
\usepackage{stfloats}
\usepackage{verbatim}
\usepackage{booktabs}
\usepackage{xcolor}  
\usepackage{cite}
\usepackage{placeins}
\usepackage{times}
\usepackage{moreverb}
\usepackage{mathrsfs}
\usepackage{bm}
\usepackage{url}
\usepackage{framed}
\usepackage{amsmath}
\usepackage{amssymb}
\usepackage[edges]{forest}

\usepackage{multicol}
\usepackage{amsmath,amssymb} 
\usepackage{times}
\usepackage{epsfig}
\usepackage{tabularx}
\usepackage{array}
\usepackage{amstext}
\usepackage{relsize}
\usepackage{lipsum}
\usepackage{color}
\usepackage{multicol}
\usepackage{longtable}
\usepackage{xtab,booktabs}
\usepackage{marginnote}

\usepackage{amsmath,amsthm,amssymb}

\usepackage{booktabs}

\usepackage{multirow}
\usepackage{enumitem}
\usepackage{makecell}
\usepackage{amsmath}
\usepackage{textcomp}
\usepackage{stfloats}
\usepackage{url}
\usepackage{verbatim}
\usepackage{tikz}
\usepackage{balance}
\usepackage[colorlinks, linkcolor=blue, citecolor=blue, urlcolor=blue]{hyperref}
\usepackage{ragged2e}
\usepackage{wrapfig}

\newcommand{\eg}{\textit{e}.\textit{g}.}

\usepackage{balance}

\definecolor{hidden-draw}{RGB}{20,68,106}
\definecolor{hidden-pink}{RGB}{234, 131, 121}
\definecolor{lightred}{RGB}{220,92,96}
\definecolor{deepblue}{RGB}{125,174,224}
\definecolor{lightpurp}{RGB}{179,149,189}
\definecolor{lightpurple}{RGB}{130, 132, 131}
\definecolor{lightgray}{gray}{0.9}

\definecolor{hiddenc1}{RGB}{59, 118, 122}
\definecolor{hiddenc2}{RGB}{69,105,144}
\definecolor{hiddenc3}{RGB}{130,130,170}

\definecolor{hid-vae}{RGB}{125,174,224}
\definecolor{hid-gnn}{RGB}{179,149,189}
\definecolor{hid-trans}{RGB}{122, 199,226}
\definecolor{hid-dm}{RGB}{225, 225, 255}
\definecolor{hid-llm}{RGB}{84,190,170}
\definecolor{hid-ssl}{RGB}{176,217,146}
\definecolor{hid-dms}{RGB}{238, 144, 59}

\newtheorem{definition}{Definition}

\usepackage[flushleft]{threeparttable}

\begin{document}

\title{A Comprehensive Survey on Benchmarks and Solutions in Software Engineering of LLM-Empowered Agentic System}

\author{
Jiale Guo$^{1,\ast}$, 
Suizhi Huang$^{2,\ast}$, 
Mei Li$^{2,4,\ast}$\thanks{$\ast$ These authors contributed equally to this work}, 
Dong Huang$^{5}$, 
Xingsheng Chen$^{5}$,\\
Regina Zhang$^{6,\dagger}$,
Zhijiang Guo$^{3}$,
Han Yu$^{2,\dagger}$, 
Siu-Ming Yiu$^{5,\dagger}$,\\
Pietro Lio$^{6}$,
Kwok-Yan Lam$^{1,\dagger}$\thanks{$\dagger$ Corresponding author}
\\
\small $^{1}$Digital Trust Centre, Nanyang Technological University, Singapore \\
\small $^{2}$College of Computing and Data Science, Nanyang Technological University, Singapore \\
\small $^{3}$The Hong Kong University of Science and Technology, Hong Kong \\
\small $^{4}$School of Computer Science, Shanghai Jiao Tong University, Shanghai, China \\
\small $^{5}$School of Computing and Data Science, The University of Hong Kong, Hong Kong\\
\small $^{6}$School of Computer Science and Technology, The University of Cambridge, UK \\
}

\markboth{Journal of \LaTeX\ Class Files,~Vol.~14, No.~8, August~2021}%
{Shell \MakeLowercase{\textit{et al.}}: A Sample Article Using IEEEtran.cls for IEEE Journals}

\IEEEpubid{0000--0000/00\$00.00~\copyright~2021 IEEE}

\maketitle

\begin{abstract}
The integration of Large Language Models (LLMs) into software engineering has driven a transition from traditional rule-based systems to autonomous agentic systems capable of solving complex problems. However, systematic progress is hindered by a lack of comprehensive understanding of how benchmarks and solutions interconnect. This survey addresses this gap by providing the first holistic analysis of LLM-powered software engineering, offering insights into evaluation methodologies and solution paradigms. We review over 150 recent papers and propose a taxonomy along two key dimensions: (1) Solutions, categorized into prompt-based, fine-tuning-based, and agent-based paradigms, and (2) Benchmarks, including tasks such as code generation, translation, and repair. Our analysis highlights the evolution from simple prompt engineering to sophisticated agentic systems incorporating capabilities like planning, reasoning, memory mechanisms, and tool augmentation. To contextualize this progress, we present a unified pipeline illustrating the workflow from task specification to deliverables, detailing how different solution paradigms address various complexity levels.
Unlike prior surveys that focus narrowly on specific aspects, this work connects 50+ benchmarks to their corresponding solution strategies, enabling researchers to identify optimal approaches for diverse evaluation criteria. We also identify critical research gaps and propose future directions, including multi-agent collaboration, self-evolving systems, and formal verification integration. This survey serves as a foundational guide for advancing LLM-driven software engineering.
We maintain a GitHub repository that continuously updates the reviewed and related papers at \url{https://github.com/lisaGuojl/LLM-Agent-SE-Survey}.
\end{abstract}


\section{Introduction}\label{sec:intro}
\IEEEPARstart{S}{oftware} Engineering aims to systematically develop high-quality, reliable, and maintainable software systems through disciplined methodologies, tools, and best practices~\cite{amazon2025qdeveloper,opendevin2024}. Key objectives include effectiveness, efficiency, maintainability and scalability~\cite{liu2025efficientagents,zhang2025economics}. And the main tasks of software engineering are code generation~\cite{ma2025rethinking}, code translation~\cite{ke2025mutual} and program repair~\cite{zheng2025patch}. Earlier studies for software engineering mainly adopt generating codes via rule-based systems and template-driven approaches, which rely on predefined patterns and heuristics. However, these traditional methods~\cite{correa2000object} often struggle with complex programming scenarios and lack the flexibility to handle diverse coding styles and requirements. Recent advances~\cite{liu2025efficientagents,zhang2025economics} have shifted towards learning-based approaches, particularly leveraging deep learning models and large language models to better capture programming patterns and semantics.

\begin{figure}[tpb!]
    \centering
    \includegraphics[width=\linewidth]{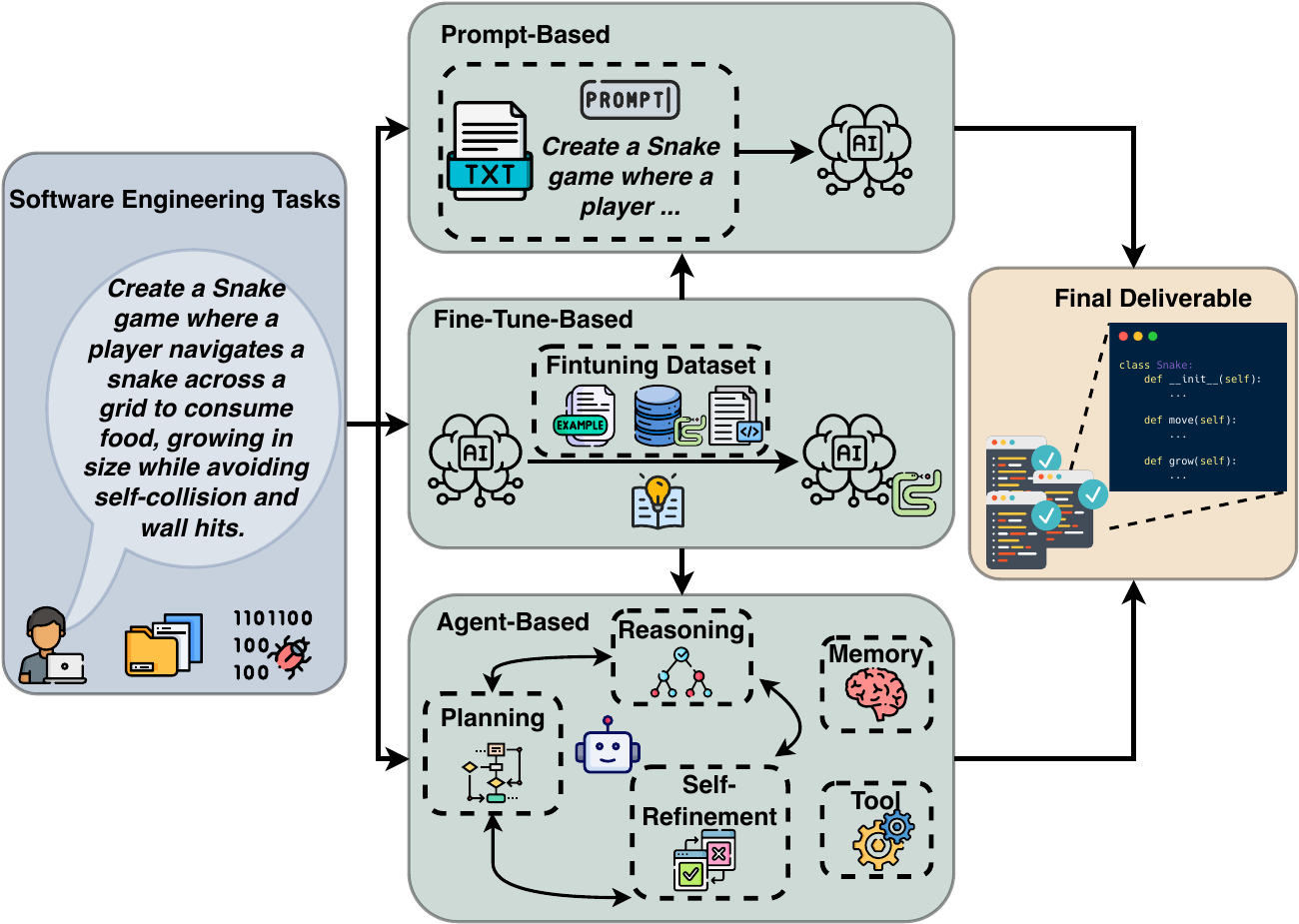}
    \vspace{-0.2in}
    \caption{Illustrations of process of LLM-empowered software engineering}
    \label{fig:intro}
    \vspace{-0.2in}
\end{figure}

\IEEEpubidadjcol

The emergence of Large Language Models (LLMs) has fundamentally transformed the landscape of software engineering. As illustrated in Figure~\ref{fig:intro}, modern LLM-empowered software engineering follows a sophisticated multi-stage process that begins with diverse software engineering tasks - from creating simple games to complex enterprise applications. Unlike traditional approaches, this paradigm employs three distinct methodological frameworks: Prompt-Based approaches that directly query LLMs with carefully crafted instructions, Fine-Tune-Based methods that adapt models to specific domains through supervised learning, and Agent-Based systems that orchestrate multiple components including planning, reasoning, memory, and tool integration. These systems demonstrate remarkable capabilities in understanding natural language specifications and generating contextually appropriate code, ultimately producing final deliverables that range from functional applications to comprehensive software solutions. The integration of these approaches enables LLM-empowered systems to autonomously handle complex programming logic across diverse languages and frameworks, making them increasingly valuable in modern software development workflows where they can operate with minimal human intervention while maintaining high-quality output standards.


\begin{table*}[t]
    \centering
    \caption{Comparison of related surveys on software engineering (PR denotes program repair. CG denotes code generation. SE denotes software engineering.)}
    \vspace{-0.05in}
    \label{tab:related_survey_transposed}
    \small
    \setlength{\tabcolsep}{1.70mm}{
    \begin{tabular}{l|cccccccc}
        \toprule
        Features              & Year & Taxonomy & Pipeline & Prompt & Agent & Benchmarks & Solutions & Scope \\
        \midrule
        Zhang et al. \cite{zhang2023survey} & 2023 & \checkmark & \checkmark & $\times$ & $\times$ & \checkmark & \checkmark & PR \\
        Jiang et al. \cite{jiang2024survey} & 2024 & \checkmark & \checkmark & \checkmark & \checkmark & $\times$ & \checkmark & CG \\
        Wang et al. \cite{wang2025ai}      & 2025 & \checkmark & \checkmark & \checkmark & \checkmark & $\times$   & \checkmark & CG \\
        Dong et al. \cite{dong2025survey}  & 2025 & \checkmark & \checkmark & \checkmark & \checkmark & $\times$   & \checkmark & CG \\
        Sapkota et al. \cite{sapkota2025vibe} & 2025 & $\times$   & $\times$   & \checkmark & \checkmark & $\times$   & \checkmark & CG \\
        \textbf{Ours}          & 2025 & \checkmark & \checkmark & \checkmark & \checkmark & \checkmark & \checkmark & PR, CG, SE \\
        \bottomrule
    \end{tabular}}
    \vspace{-0.15in}
\end{table*}

\textbf{Existing surveys:} With the success of LLM-driven methods for software engineering, many researchers have begun to summarize various aspects of this rapidly evolving field. As shown in Table~\ref{tab:related_survey_transposed}, existing surveys exhibit distinct limitations in their scope and coverage. Zhang et al.\cite{zhang2023survey} (2023) provide an early comprehensive study but focus exclusively on program repair without covering agent-based approaches or prompt engineering techniques. While recent surveys from 2024-2025\cite{jiang2024survey,wang2025ai,dong2025survey} have made significant progress in documenting agent capabilities - including planning and decomposition, reasoning and self-refinement, tool augmentation, and memory mechanisms - they notably lack coverage of benchmarks, making it difficult for researchers to evaluate and compare different approaches systematically. Sapkota et al.\cite{sapkota2025vibe} (2025) provide insights into agent components and prompting strategies but omit both taxonomy and pipeline perspectives, limiting the structural understanding of the field. Most critically, all existing surveys focus on narrow scopes: they either address only code generation\cite{jiang2024survey,wang2025ai,dong2025survey,sapkota2025vibe} or program repair~\cite{zhang2023survey} in isolation, without providing a holistic view of software engineering tasks. None of these surveys successfully bridge the gap between benchmarks and solutions---a crucial connection needed for advancing the field.

\textbf{Motivations of This Survey:} The limitations identified in existing surveys reveal critical gaps that motivate our comprehensive study. \textbf{First}, while recent surveys~\cite{jiang2024survey,wang2025ai,dong2025survey} document individual agent capabilities, they lack a systematic taxonomy that captures how these components integrate into complete agentic systems, from architectural design to problem-solving strategies. \textbf{Second}, the absence of benchmark coverage in most recent surveys means researchers cannot effectively evaluate the agent capabilities being proposed, creating a disconnect between theoretical advances and practical validation. \textbf{Third}, the narrow focus on either code generation or program repair prevents understanding of how LLM-empowered systems can address the full spectrum of software engineering challenges, including code translation, testing, and documentation. \textbf{Fourth}, without connecting benchmarks to solutions, researchers struggle to identify which approaches work best for specific tasks and where improvements are needed. \textbf{Finally}, the rapid evolution of this field, particularly in areas such as multi-agent collaboration, self-refinement mechanisms, and advanced tool integration strategies, necessitates an up-to-date survey that captures these recent breakthroughs. Our survey addresses all these gaps by providing comprehensive coverage across both benchmarks and solutions, spanning all major software engineering tasks, and offering systematic organization through taxonomy and pipeline perspectives that enable researchers to understand not only what solutions exist, but also how to evaluate them effectively.

\textbf{Selection Papers of This Survey:} To ensure comprehensive coverage and high quality, we systematically collected recent papers from multiple sources including top-tier conferences (NeurIPS, ICML, ICLR, ACL, EMNLP, ICSE, FSE, ASE), journals (TSE, TOSEM, TPAMI), and a small amount of preprint servers (arXiv, OpenReview). Our selection criteria include: (1) Papers published between 2023 and 2025 that explicitly address LLM-based approaches for software engineering; (2) Works that contribute novel benchmarks, datasets, or evaluation metrics; (3) Studies proposing new methods, architectures, or techniques for LLM-empowered agentic systems; (4) Papers with significant empirical evaluation demonstrating practical impact. 

To provide comprehensive understanding on software engineering, we provide a taxonomy on LLM-powered agentic system, shown in Figure~\ref{fig:taxonomy}. Our taxonomy is organized into two major dimensions: Solutions and Benchmarks. We also provide a pipeline to illustrate the whole process of LLM-empowered software from raw data to applications. To summarize, we provide our contributions as follows:

\begin{itemize}
\item \textbf{Comprehensive taxonomy bridging benchmarks and solutions:} We provide the first comprehensive taxonomy that systematically organizes 150+ recent papers in LLM-empowered software engineering, uniquely connecting evaluation benchmarks with their corresponding solution approaches. Unlike existing surveys that treat these aspects separately, our framework categorizes solutions into prompt-based, fine-tuning-based, and agent-based paradigms while simultaneously mapping them to relevant benchmarks across code generation, translation, repair, and other tasks, enabling researchers to identify which methods work best for specific evaluation criteria.

\item \textbf{Unified pipeline with full-spectrum coverage:} We present a unified pipeline that illustrates the complete workflow of LLM-empowered software engineering from initial task specification to final deliverables. Our analysis is the first to provide holistic coverage of all major software engineering tasks rather than focusing on isolated aspects like code generation or program repair. We detail how agent capabilities (planning and decomposition, reasoning and self-refinement, memory mechanisms, and tool augmentation) work synergistically throughout this pipeline to solve complex engineering challenges.

\begin{figure*}[t]
    \centering
    \includegraphics[width=0.9\linewidth]{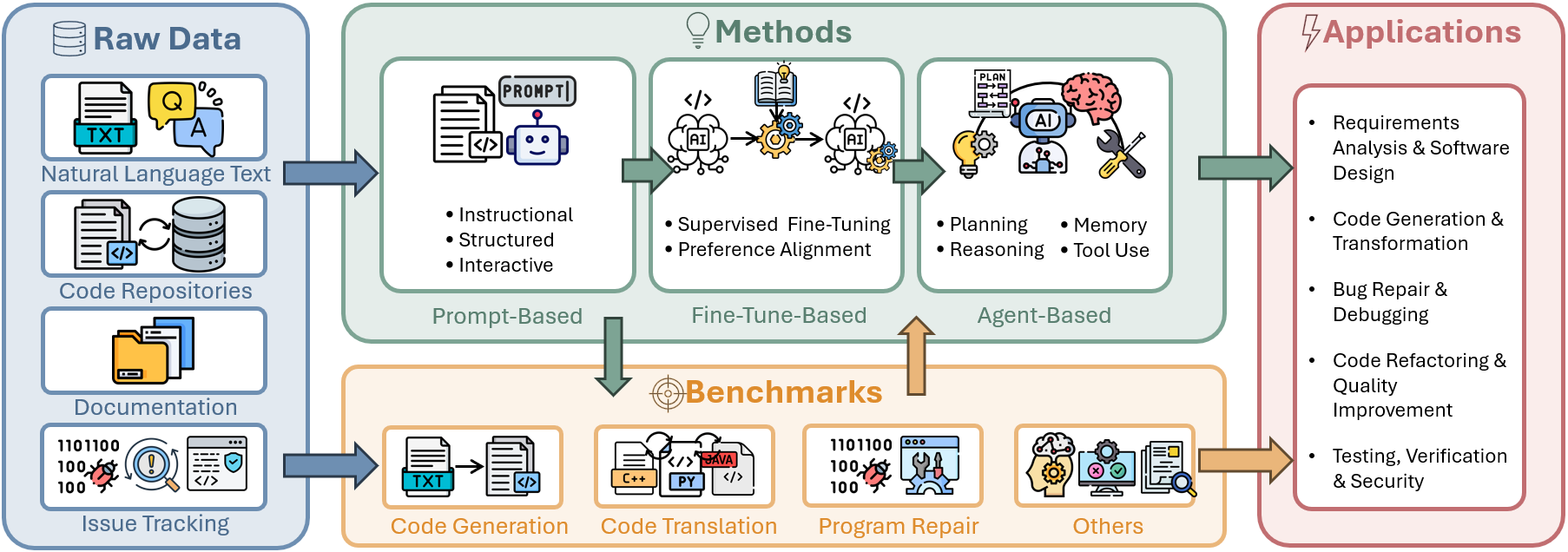}
    \caption{Pipeline of this survey.}
    \label{fig:pipeline}
    \vspace{-0.2in}
\end{figure*}

\item \textbf{Actionable insights and emerging research directions:} Based on our systematic analysis of the field's current state, we identify critical research gaps and provide concrete future directions including multi-agent collaboration strategies for complex software projects, self-evolving code generation systems with continuous learning capabilities, cross-domain knowledge transfer mechanisms, and integration of formal verification methods with LLM-based approaches. These insights offer researchers clear pathways to advance beyond current limitations and develop next-generation software engineering systems.
\end{itemize}

\section{Preliminaries and Problem Definition}
\label{sec:preliminaries}

In this section, we provide several key benchmark task definitions of software engineering, namely code generation, code translation, and program repair. The detailed illustrations are shown as follows:

\begin{definition}[Code Generation]
We transform high-level concepts into executable codes. We provide formula definitions, which is shown as $\Gamma = \mathcal{S} \times \mathcal{C} \rightarrow \mathcal{P}$, where $\mathcal{S}$ encompasses natural language descriptions, pseudocode, and input-output examples, while $\mathcal{C}$ represents contextual constraints including available APIs, libraries, and environmental factors, and $\mathcal{P}$ denotes the space of executable programs. Traditional approaches to this mapping include rule-based templates ($\Gamma_{\text{rule}}$) and statistical methods ($\Gamma_{\text{stat}}$). 
\end{definition}

\begin{definition}[Code Translation]
We convert source code from a source language to a target language while preserving functionality. We define this process with the mapping $\Gamma = \mathcal{S} \times \mathcal{T} \times \mathcal{C} \rightarrow \mathcal{P}{\mathcal{T}}$, where $\mathcal{S}$ represents the source code in the original language, $\mathcal{T}$ denotes the target programming language, and $\mathcal{C}$ captures contextual constraints such as libraries, APIs, and platform requirements. The output $\mathcal{P}{\mathcal{T}}$ is the translated code in the target language that is functionally equivalent to the source. Traditional approaches include rule-based systems ($\Gamma_{\text{rule}}$) and intermediate representation-based methods ($\Gamma_{\text{IR}}$).
\end{definition}

\begin{definition}[Program Repair]
We transform a faulty program into a repaired version that meets given specifications. The repair process is defined by $\Gamma = \mathcal{P}_f \times \mathcal{S} \times \mathcal{C} \rightarrow \mathcal{P}_r$, where $\mathcal{P}f$ is the faulty program, $\mathcal{S}$ represents specifications (\eg, test suites, assertions), $\mathcal{C}$ captures contextual constraints (\eg, time, resource, and code quality requirements), and $\mathcal{P}r$ is the repaired program that satisfies $\mathcal{S}$. Traditional approaches include generate-and-validate methods ($\Gamma{\text{gen-val}}$) and constraint-based repair ($\Gamma{\text{constr}}$).
\end{definition}

\tikzstyle{my-box}=[
    rectangle,
    draw=hidden-draw,
    rounded corners,
    align=center,
    text opacity=1,
    minimum height=1.5em,
    minimum width=5em,
    inner sep=2pt,
    fill opacity=.8,
    line width=0.8pt,
]

\tikzstyle{leaf-head}=[my-box, minimum height=1.5em,
    draw=black, 
    text=black, font=\normalsize,
    inner xsep=2pt,
    inner ysep=4pt,
    line width=0.8pt,
]

\tikzstyle{leaf-task}=[my-box, minimum height=2.5em,
    draw=black, 
    text=black, font=\normalsize,
    inner xsep=2pt,
    inner ysep=4pt,
    line width=0.8pt,
]
\tikzstyle{leaf-taska}=[my-box, minimum height=2.5em,
    draw=black, 
    text=black, font=\normalsize,
    inner xsep=2pt,
    inner ysep=4pt,
    line width=0.8pt,
]
\tikzstyle{leaf-task1}=[my-box, minimum height=2.5em,
    draw=black, 
    text=black, font=\normalsize,
    inner xsep=2pt,
    inner ysep=4pt,
    line width=0.8pt,
]
\tikzstyle{modelnode-task1}=[my-box, minimum height=1.5em,
    draw=black, 
    text=black, font=\normalsize,
    inner xsep=2pt,
    inner ysep=4pt,
    line width=0.8pt,
]
\tikzstyle{leaf-task2}=[my-box, minimum height=2.5em,
    draw=black, 
    text=black, font=\normalsize,
    inner xsep=2pt,
    inner ysep=4pt,
    line width=0.8pt,
]
\tikzstyle{modelnode-task2}=[my-box, minimum height=1.5em,
    draw=black, 
    text=black, font=\normalsize,
    inner xsep=2pt,
    inner ysep=4pt,
    line width=0.8pt,
]
\tikzstyle{leaf-task3}=[my-box, minimum height=2.5em,
    draw=black, 
    text=black, font=\normalsize,
    inner xsep=2pt,
    inner ysep=4pt,
    line width=0.8pt,
]
\tikzstyle{modelnode-task3}=[my-box, minimum height=1.5em,
    draw=black, 
    text=black, font=\normalsize,
    inner xsep=2pt,
    inner ysep=4pt,
    line width=0.8pt,
]

\tikzstyle{leaf-task4}=[my-box, minimum height=2.5em,
    draw=black, 
    text=black, font=\normalsize,
    inner xsep=2pt,
    inner ysep=4pt,
    line width=0.8pt,
]
\tikzstyle{modelnode-task4}=[my-box, minimum height=1.5em,
    draw=black, 
    text=black, font=\normalsize,
    inner xsep=2pt,
    inner ysep=4pt,
    line width=0.8pt,
]

\tikzstyle{leaf-task5}=[my-box, minimum height=2.5em,
    draw=black, 
    text=black, font=\normalsize,
    inner xsep=2pt,
    inner ysep=4pt,
    line width=0.8pt,
]

\tikzstyle{modelnode-task5}=[my-box, minimum height=1.5em,
    draw=black, 
    text=black, font=\normalsize,
    inner xsep=2pt,
    inner ysep=4pt,
    line width=0.8pt,
]

\tikzstyle{leaf-task6}=[my-box, minimum height=2.5em,
    draw=black, 
    text=black, font=\normalsize,
    inner xsep=2pt,
    inner ysep=4pt,
    line width=0.8pt,
]

\tikzstyle{leaf-task7}=[my-box, minimum height=2.5em,
    draw=black, 
    text=black, font=\normalsize,
    inner xsep=2pt,
    inner ysep=4pt,
    line width=0.8pt,
]
\tikzstyle{leaf-task8}=[my-box, minimum height=2.5em,
    draw=black, 
    text=black, font=\normalsize,
    inner xsep=2pt,
    inner ysep=4pt,
    line width=0.8pt,
]

\tikzstyle{leaf-task9}=[my-box, minimum height=2.5em,
    draw=black, 
    text=black, font=\normalsize,
    inner xsep=2pt,
    inner ysep=4pt,
    line width=0.8pt,
]
\tikzstyle{leaf-task10}=[my-box, minimum height=2.5em,
    draw=black, 
    text=black, font=\normalsize,
    inner xsep=2pt,
    inner ysep=4pt,
    line width=0.8pt,
]

\tikzstyle{modelnode-task6}=[my-box, minimum height=1.5em,
    draw=black, 
    text=black, font=\normalsize,
    inner xsep=2pt,
    inner ysep=4pt,
    line width=0.8pt,
]

\tikzstyle{modelnode-task7}=[my-box, minimum height=1.5em,
    draw=black, 
    text=black, font=\normalsize,
    inner xsep=2pt,
    inner ysep=4pt,
    line width=0.8pt,
]
\tikzstyle{modelnode-task8}=[my-box, minimum height=1.5em,
    draw=black, 
    text=black, font=\normalsize,
    inner xsep=2pt,
    inner ysep=4pt,
    line width=0.8pt,
]
\tikzstyle{modelnode-task9}=[my-box, minimum height=1.5em,
    draw=black, 
    text=black, font=\normalsize,
    inner xsep=2pt,
    inner ysep=4pt,
    line width=0.8pt,
]

\tikzstyle{leaf-paradigms}=[my-box, minimum height=2.5em,
    draw=black, 
    text=black, font=\normalsize,
    inner xsep=2pt,
    inner ysep=4pt,
    line width=0.8pt,
]
\tikzstyle{leaf-others}=[my-box, minimum height=2.5em,
    draw=black, 
    text=black, font=\normalsize,
    inner xsep=2pt,
    inner ysep=4pt,
    line width=0.8pt,
]
\tikzstyle{leaf-other}=[my-box, minimum height=2.5em,
    draw=orange!80, 
    fill=orange!15,  
    text=black, font=\normalsize,
    inner xsep=2pt,
    inner ysep=4pt,
    line width=0.8pt,
]

\tikzstyle{modelnode-task}=[my-box, minimum height=1.5em,
    draw=black, 
    text=black, font=\normalsize,
    inner xsep=2pt,
    inner ysep=4pt,
    line width=0.8pt,
]

\tikzstyle{modelnode-paradigms}=[my-box, minimum height=1.5em,
    draw=black, 
    text=black, font=\normalsize,
    inner xsep=2pt,
    inner ysep=4pt,
    line width=0.8pt,
]
\tikzstyle{modelnode-others}=[my-box, minimum height=1.5em,
    draw=black, 
    text=black, font=\normalsize,
    inner xsep=2pt,
    inner ysep=4pt,
    line width=0.8pt,
]
\tikzstyle{modelnode-other}=[my-box, minimum height=1.5em,
    draw=black, 
    text=black, font=\normalsize,
    inner xsep=2pt,
    inner ysep=4pt,
    line width=0.8pt,
]
\begin{figure*}[!ht]
    \centering
    \resizebox{1\textwidth}{!}{
        \begin{forest}
            forked edges,
            for tree={
                grow=east,
                reversed=true,
                anchor=base west,
                parent anchor=east,
                child anchor=west,
                base=left,
                font=\normalsize,
                rectangle,
                draw=hidden-draw,
                rounded corners,
                align=center,
                minimum width=1em,
                edge+={darkgray, line width=1pt},
                s sep=3pt,
                inner xsep=0pt,
                inner ysep=3pt,
                line width=0.8pt,
                ver/.style={rotate=90, child anchor=north, parent anchor=south, anchor=center},
            }, 
            [
                Benchmarks and Solutions in Software Engineering,leaf-head, ver
                [
                    Solutions\\(\textbf{Sec.~\ref{sec:solution}}), leaf-paradigms,text width=8em
                    [
                        Prompt-Based\\(\textbf{Sec.~\ref{sec:promt}}), leaf-paradigms, text width=9.5em
                        [
                            Instructional, leaf-paradigms, text width=10.5em
                            [
                            ~TGen~\cite{mathews2024test}{,}~\cite{fu2025first}{,}PACGBI~\cite{sarschar2024pacgbi}{,}~\cite{pandini2025exploratory}{,}~\cite{kc2025demystifying}{,} AdaptiveLLM~\cite{cheng2025adaptivellm}{,}~\cite{liurevisiting}{,}\\Code-PLMs~\cite{zhao2024current}{,}~\cite{kondo2024improving}, text width = 31em
                            ]
                        ]    
                        [
                            Structured, leaf-paradigms, text width=10.5em
                            [
                            ~SynFix~\cite{tang2025synfix}{,}LLM4TDG~\cite{liu2025llm4tdg}{,}Codes\cite{zan2024codes}{,}~\cite{peng2025can}  , text width = 31em
                            ]
                        ]          
                        [
                            Interactive, leaf-paradigms, text width=10.5em
                            [
                            ~Clarigen~\cite{miao2025clarigen}{,}MACEDON~\cite{liu2025macedon}{,}ChatDBG\cite{levin2025chatdbg}{,}~\cite{fu2025first}  , text width = 31em
                            ]
                        ]    
                    ]
                    [
                        Fine-Tune-Based\\(\textbf{Sec.~\ref{sec:finetune}}), leaf-paradigms, text width=9.5em
                        [Supervised Fine-Tuning, leaf-paradigms, text width=10.5em
                            [
                            SWE-GPT~\cite{ma2025swe}{,}D3~\cite{piterbargd3}{,}SWE-Gym\cite{pan2024training}{,}
                            SWE-Dev~\cite{wang2025swe}{,} PoPilot~\cite{zhang2025building}{,}\\~\cite{cassano2023can}{,}MORepair~\cite{yang2024morepair}{,}BLAZE~\cite{chakraborty2025blaze}{,}~\cite{sagtani2025improving}{,}~\cite{shypula2025automated}{,}
                            SODA~\cite{chen2025smaller}{,}PtTrust~\cite{huang2025risk}{,}\\CGM~\cite{tao2025code}{,}~\cite{wang2025code}{,}Co-PatcheR~\cite{tang2025co}{,}~\cite{sapronov2025pretraining}, text width = 31em
                            ]
                        ]             
                        [Preference Alignment, leaf-paradigms, text width=10.5em
                            [
                            RL-Based, leaf-paradigms, text width=9.5em
                                [~ReST-MCTS*~\cite{zhang2024rest}{,}SEAlign~\cite{zhang2025sealign}{,}
                                ORPS~\cite{yu2024reasoning}{,}\\Kimi K2~\cite{team2025kimi}{,}CBR~\cite{guo2025optimizing}, text width = 19.9em]
                            ]
                            [
                            RL-Free, leaf-paradigms, text width=9.5em
                                [~SelfCodeAlign~\cite{wei2024selfcodealign}{,}LPO~\cite{saqib2025teaching}, text width = 19.9em]
                            ]
                        ]       
                    ]
                    [
                        Agent-Based\\(\textbf{Sec.~\ref{sec:agent}}), leaf-paradigms, text width=9.5em
                        [
                        Planning \& \\Decomposition, leaf-paradigms, text width=10.5em
                             [MAGIS~\cite{tao2024magis}{,}AgileCoder~\cite{nguyen2025agilecoder}{,}MASAI~\cite{arora2024masai}{,}Artemis~\cite{giavrimis2025artemis}{,}
                             XpandA~\cite{xiao2025long}{,}\\
                             Co-PatcheR~\cite{tang2025co}{,}Codes~\cite{zan2024codes}{,}LLMDFA~\cite{wang2024llmdfa}{,}DRCodePilot~\cite{zhao2024enhancing}{,}\\MarsCode~\cite{liu2024marscode}{,}AIOps~\cite{shetty2024building}{,}ExecutionAgent~\cite{bouzenia2025you}{,}ExploraCoder~\cite{wang2024exploracoder}{,}\\~LingmaAgent~\cite{ma2025improving}{,}RepoUnderstander~\cite{ma2024understand}{,}Alphaverus~\cite{aggarwal2024alphaverus}{,}Agentless~\cite{xia2025demystifying}{,}\\PatchPilot~\cite{li2025patchpilot}{,}AGDebugger~\cite{epperson2025interactive}, text width=31em]
                        ]
                        [
                        Reasoning \& \\Self-Refinement, leaf-paradigms, text width=10.5em
                        [~AutoCodeRover~\cite{zhang2024autocoderover}{,}PATCHAGENT~\cite{zheng2025patch}{,}SWE-Dev~\cite{wang2025swe}{,}Repo2Run~\cite{hu2025llm}{,}\\MarsCode~\cite{liu2024marscode}{,}~\cite{wang2025solved}{,}Alphaverus~\cite{aggarwal2024alphaverus}{,}HULA~\cite{takerngsaksiri2025human}{,}~\cite{omidvar2024evaluating}{,}ROSE~\cite{reiss2025rose}{,}\\STOP~\cite{zelikman2024self}{,}~\cite{dou2024s}{,}Artemis~\cite{giavrimis2025artemis}{,}Nemotron-CORTEXA~\cite{sohrabizadehnemotron}{,}\\Co-PatcheR~\cite{tang2025co}{,}~\cite{nouri2025simulation}{,}Codellaborator~\cite{pu2025assistance}{,}LLM4TDG~\cite{liu2025llm4tdg}{,}CodeV~\cite{zhang2024codev}, text width=31em
                            ]
                        ]
                        [
                        Memory Mechanisms, leaf-paradigms, text width=10.5em
                            [~LingmaAgent~\cite{ma2025alibaba}{,}RepoUnderstander~\cite{ma2024understand}{,}CGM~\cite{tao2025code}{,}COLLMS~\cite{truong2024coordinating}{,}~\cite{ehsani2025towards}{,}\\AutoCodeRover~\cite{zhang2024autocoderover}{,}BLAZE~\cite{chakraborty2025blaze}{,}AEGIS~\cite{wang2025aegis}{,}EvoR~\cite{su2024evor}{,}XpandA~\cite{xiao2025long}  , text width=31em]
                        ]
                        [
                        Tool Augmentation, leaf-paradigms, text width=10.5em
                            [
                            ~SWE-agent~\cite{yang2024swe}{,}ExecutionAgent~\cite{bouzenia2025you}{,}Repo2Run~\cite{hu2025llm}{,}MarsCode~\cite{liu2024marscode}{,}\\Mediator~\cite{li2025enhancing}{,}AutoCodeRover~\cite{zhang2024autocoderover}{,}LLMDFA~\cite{wang2024llmdfa}{,}AgileCoder~\cite{nguyen2025agilecoder}{,}\\BLAST~\cite{kitsios2025automated}{,}Alphaverus~\cite{aggarwal2024alphaverus}{,}PatchPilot~\cite{li2025patchpilot}{,}~\cite{nouri2025simulation}{,}CodeV~\cite{zhang2024codev}{,}\\OpenHands-Versa~\cite{soni2025coding}{,}SWE-Gym~\cite{pan2024training}{,}Kimi K2~\cite{team2025kimi}, text width=31em
                            ]
                        ]
                    ]
                ]
                [
                     Benchmarks\\(\textbf{Sec.~\ref{sec:benchmarks}}), leaf-task,text width=8em
                    [
                        Code Generation\\(\textbf{Sec.~\ref{sec:code_generation}}), leaf-task10, text width=9.5em
                        [~HumanEval~\cite{chen2021evaluating}{, } MBPP~\cite{austin2021program}{, }CodeXGLUE~\cite{lu2021codexglue}{, }CodeNet~\cite{puri2021codenet}{, }HumanEval+~\cite{Liu2023is}{, } McEval~\cite{chai2025mceval}{, }\\HumanEval Pro{, }MBPP Pro~\cite{yu2025humanevalpro}{, }CodeScope~\cite{yan2024codescope}{, }BigCodeBench~\cite{zhuo2025bigcodebench}{, } ClassEval~\cite{du2024classeval}{, }\\SAFIM~\cite{gong2024evaluation}{, } EffiBench-X~\cite{qing2025effibench}{,}CVLEVER~\cite{thakur2025clever}{,}SWEE-Bench~\cite{vergopoulos2025automated}{, }DevEval~\cite{li2024deveval}{, }\\ RepoCOD~\cite{liang2025repocod}{, }HumanEvo~\cite{Zheng2025humanevo}{, }R2E-Eval~\cite{jain2024r2e}{, }SWE-smith~\cite{yang2025swesmith}{,}SWE-rebench~\cite{badertdinov2025swe}{,}\\DyCodeEval~\cite{chen2025dycodeeval}{, }HumanEvalComm~\cite{wu2025humanevalcomm}{, } InterCode~\cite{fu2025first}{, }MaintainCoder~\cite{wang2025maintaincoder}{,}\\HumanEval-X~\cite{zheng2023codegeex}{,}XCodeEval~\cite{khan2024xcodeeval}{,}CodeCrossBench~\cite{nie2023crosscodebench}{,}SECCODEPLT~\cite{yang2024seccodeplt}, text width=43.1em]
                    ]
                    [
                        Code Translation\\(\textbf{Sec.~\ref{sec:code_gtranslation}}), leaf-taska, text width=9.5em
                            [ CodeXGLUE~\cite{lu2021codexglue}{, }CodeNet~\cite{puri2021codenet}{, }AVATAR~\cite{ahmad2023avatar}{, }CodeTransOcean~\cite{yan2023codetransocean}{,} HumanEval-X~\cite{zheng2023codegeex}{,}\\ XCodeEval~\cite{khan2024xcodeeval}{,}ClassEval-T~\cite{du2024evaluating}{,}CRUST-Bench~\cite{khatry2025crustbench}{,}RustRepoTrans~\cite{ou2024repository}{,}RepoTransBench~\cite{wang2024repotransbench}{,}\\~AlphaTrans~\cite{ibrahimzada2025alphatrans}{,}F2STRANS~\cite{zhang2025functiontostyle}{,}CodeEditorBench~\cite{guo2025codeeditorbench}{,}G-TransEval~\cite{jiao2023on}{,}CodeCrossBench~\cite{nie2023crosscodebench}{,}\\PolyHumanEval~\cite{tao2024unraveling}, text width=43.1em
                            ]
                    ]
                    [
                        Program Repair\\(\textbf{Sec.~\ref{sec:program_repair}}), leaf-task6, text width=9.5em
                         [ CodeXGLUE~\cite{lu2021codexglue}{, }CodeNet~\cite{puri2021codenet}{,}CodeEditorBench~\cite{guo2025codeeditorbench}{,}DebugBench~\cite{tian2024debugbench}{,}SWE-Bench~\cite{jimenez2024swebench}{, }\\RepoBugs~\cite{chen2024when}{,}RepoDebug~\cite{liu2025repodebug}{,}GHRB~\cite{Lee2024the}{,}LongCodeBench~\cite{rando2025longcodebench}{,}Breakpoint~\cite{hariharan2025breakpoint}{,}\\~Multi-SWE-bench~\cite{zan2025multiswebench}{,}OmniGIRL~\cite{guo2025omnigirl}{,}CVE-Bench~\cite{wang2025cve}{,}SEC-Bench~\cite{lee2025sec}{,}SECCODEPLT~\cite{yang2024seccodeplt}, text width=43.1em]
                    ]
                    [
                        Others\\(\textbf{Sec.~\ref{sec:others}}), leaf-taska, text width=9.5em
                            [
                            ~CRUXEval~\cite{gu2024cruxeval}{,}CRUXEval-X~\cite{xu2025cruxeval}{,}CodeJudge-Eval~\cite{zhao2025codejudge}{,}TF-Bench~\cite{he2025tf}{,}CoRe~\cite{xie2025core}{,}\\SWT-bench~\cite{mundler2024swtbench}{,}TestEval~\cite{wang2025testeval}{,}TestGenEval~\cite{jain2025testgeneval}{,}RefactorBench~\cite{gautam2025refactorbench}{,}GSO~\cite{shetty2025gso}{,}\\DI-Bench~\cite{zhang2025di}{,}Installamatic~\cite{milliken2025beyond}{,}SynBench~\cite{guo2025syncmind}, text width=43.1em
                        ]                            
                    ]
                ]
            ]
        \end{forest}
    }
    \caption{Taxonomy of existing studies for Software Engineering}
    \label{fig:taxonomy}
\end{figure*}
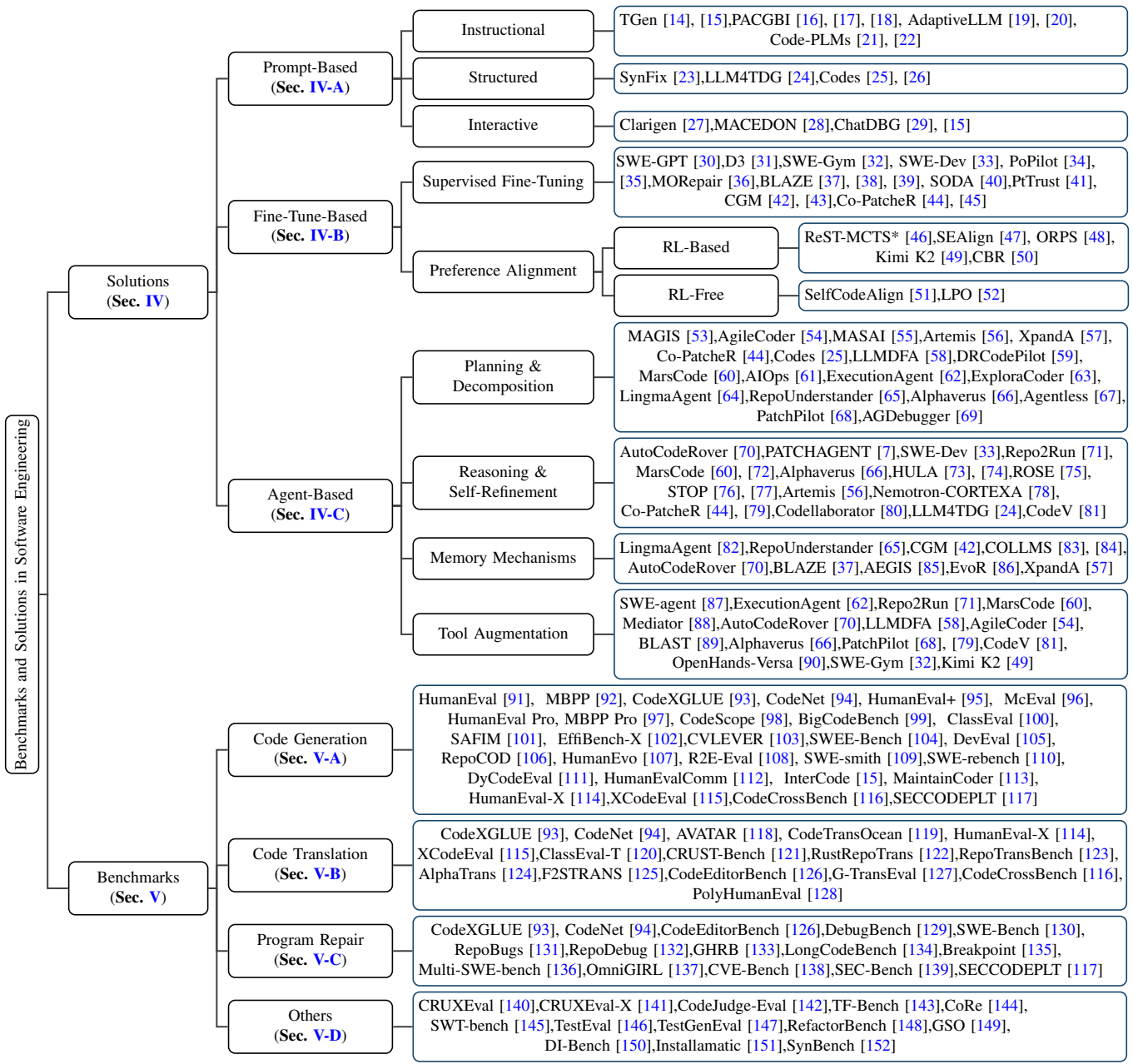
\section{Pipeline and Taxonomy} \label{sec:tax}

The application of large language models in software engineering has developed rapidly in recent years, with the emergence of various researches on advanced LLM technologies, as well as numerous benchmarks for evaluating model performance in different tasks. To better understand this rapidly evolving landscape, a systematic framework is needed to analyze and compare existing approaches. Therefore, we propose a taxonomy as a core tool for systematic reviews. 

\begin{figure}[t]
    \centering
    \includegraphics[width=0.9\linewidth]{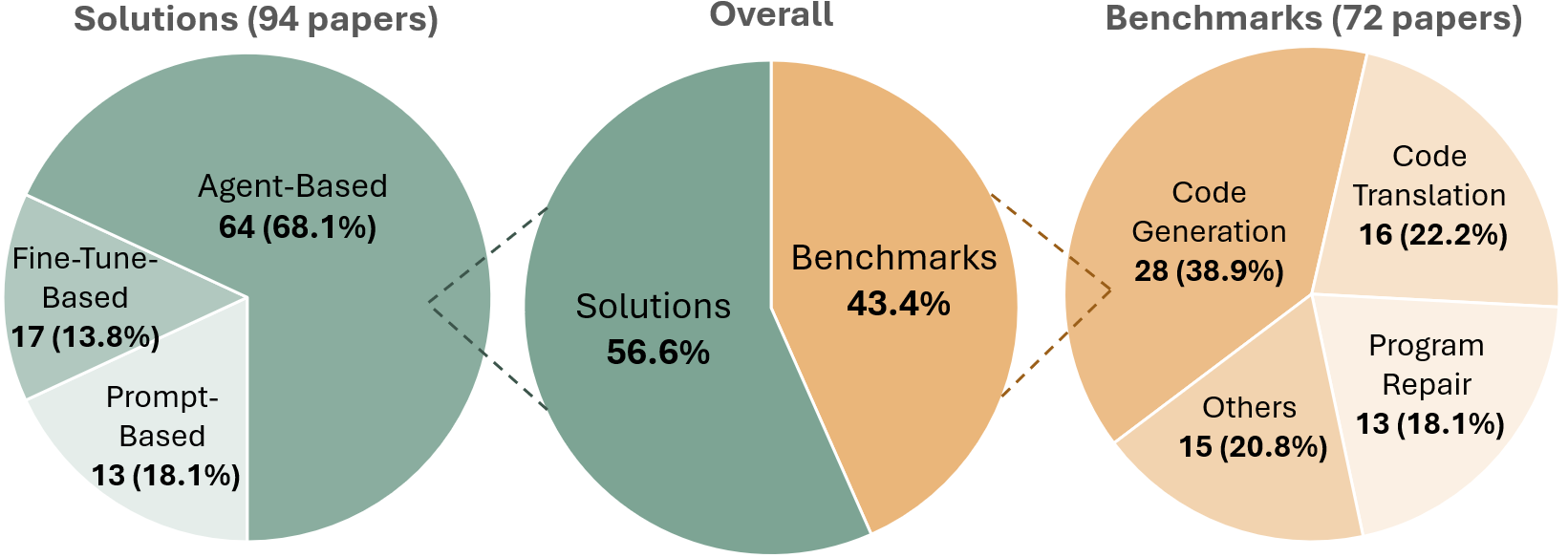}
    \caption{Overview of reviewed studies.}
    \label{fig:pie}
    \vspace{-0.2in}
\end{figure}

Distinguishing itself from existing surveys, this work reviews LLMs researches in software engineering from two dimensions: solutions and benchmarks. 
\begin{itemize}
\item Solutions: Based on the core technologies employed, the existing approaches can be categorized into three categories: (1) \textbf{Prompt-based solutions}, including instructional, structured, and interactive prompting, which are the most straightforward use of LLMs; (2) \textbf{Fine-tune-based} solutions, which enhance model performance through supervised fine-tuning as well as preference alignment via reinforcement learning (RL) or RL-free method; (3) \textbf{agent-based} solutions, which integrate task decomposition and planning, reasoning and self-improvement, memory mechanisms, and tool usage, enabling higher levels of automation and collaboration.
\item Benchmarks: Based on the targeted software engineering tasks, exist benchmarks are categorized into four types: benchmarks for (1) \textbf{code generation}; (2) \textbf{code translation}; (3) \textbf{program repair}; (4) \textbf{Others}, such as code reasoning, test generation, code refactoring, and so on.
\end{itemize}
As shown in Fig~\ref{fig:pie}, we reviewed 140 distinct papers categorized into \textbf{Solutions} (94 papers) and \textbf{Benchmarks} (72 papers), with the distribution shown for each subcategory. Papers addressing multiple aspects may appear in more than one category.
Figure \ref{fig:taxonomy} illustrates the details of this taxonomy. Leveraging this framework, we trace the paradigm shift from prompt-based methods to agent-based systems and position benchmarks according to their coverage of different software engineering scenarios, providing insights into real-world application development. 

The overall pipeline of this survey is presented in Figure~\ref{fig:pipeline}. Specifically, Section~\ref{sec:solution} reviews LLM-based solutions for software engineering, Section~\ref{sec:benchmarks} examines benchmark studies, and Section~\ref{sec:applications} discusses real-world applications in detail.

\section{Solutions}

\label{sec:solution}
LLMs have catalyzed a paradigm shift in automated software engineering, solving challenges in code generation, maintenance, and analysis. This section reviews emerging methodologies, categorized by increasing autonomy: \textbf{Prompt-based} solutions, relying on human instruction; \textbf{Fine-tune-based} solutions, adapting models to the software engineering domain; and \textbf{Agent-based} solutions, autonomous systems that plan, act, and learn. This progression shows the evolution from LLMs as tools to autonomous partners in the software development lifecycle.

\subsection{Prompt-based Solutions}
\label{sec:promt}

\begin{figure*}[th]
    \centering
    \includegraphics[width=\linewidth]{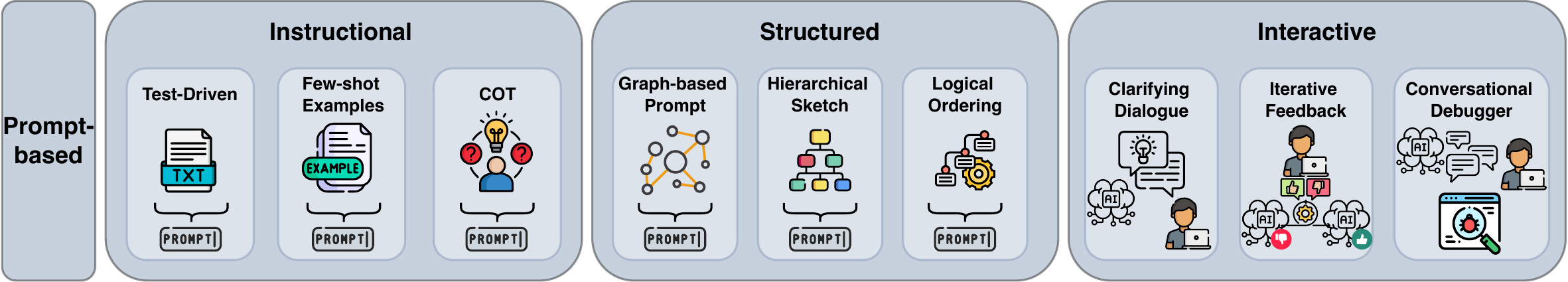}
    \caption{The components of prompt-based solutions.}
    \label{fig:prompt}
    \vspace{-0.2in}
\end{figure*}

Prompt-based methods are a direct way to interact with LLMs. The model's behavior is guided by the context and instructions in a single input, or ``prompt," without modifying its parameters. The effectiveness of these methods hinges on prompt engineering, which crafts inputs to elicit the desired output. This section explores software engineering strategies, from natural language instructions to structured and interactive prompting techniques that enhance performance and reliability, as shown in Fig.~\ref{fig:prompt}.

\subsubsection{Instructional}
Instructional prompting is the most common interaction, where a user provides a natural language task description. This approach leverages the LLM's ability to understand and follow commands, making it versatile for tasks like code generation, documentation, and bug explanation. This subsection reviews studies on instructional prompts, exploring how instruction composition influences the quality and correctness of generated code, including using examples (few-shot), test cases, or specific formatting.

A key finding is that the precision and verifiability of instructions are paramount. A powerful demonstration is using Test-Driven Development (TDD) as a prompting paradigm. Mathews \textit{et al.} found that providing an LLM with human-written unit tests alongside a natural language description serves as a detailed, executable specification, leading to a higher success rate in generating correct code~\cite{mathews2024test}. This approach provides a formal ``contract" the generated code must fulfill, reducing ambiguity.

In contrast, instructions based on less formal specifications can be less effective. Fu \textit{et al.} evaluated LLMs on iterative code generation from input-output (I/O) examples, finding model performance decreases significantly compared to when requirements are described in natural language~\cite{fu2025first}. This highlights a challenge for LLMs in abstracting complex logic from raw examples without semantic guidance from natural language.

Instructional prompts are explored across software engineering activities. For low-complexity tasks, zero-shot prompting shows promise. Sarschar \textit{et al.}'s PACGBI generates code from product backlog items, demonstrating feasibility but noting challenges with less detailed descriptions~\cite{sarschar2024pacgbi}. Prompts are also used for maintenance tasks like refactoring architectural smells \cite{pandini2025exploratory} and improving requirements engineering by analyzing feature requests\cite{kc2025demystifying}.

Prompt structure is actively researched. Liu \textit{et al.} found that with Chain-of-Thought (CoT) prompting, it is more effective to generate code first, then reasoning~\cite{liurevisiting}. Cheng \textit{et al.}'s AdaptiveLLM framework uses CoT length as a proxy for problem difficulty to select a cost-efficient LLM~\cite{cheng2025adaptivellm}.

Research has highlighted limitations of current models on specialized or less-common programming languages. Zhao \textit{et al.} evaluated Code-PLMs on the R programming language and found performance degrades due to R's unique characteristics, indicating general-purpose instructional prompting techniques do not always transfer effectively across language paradigms~\cite{zhao2024current}. To improve context for downstream tasks, Kondo \textit{et al.} propose using an LLM to convert code snippets into natural language, allowing more semantically rich searches over textual representations rather than raw code~\cite{kondo2024improving}.

\subsubsection{Structured}
To overcome natural language ambiguity, structured prompting imposes a formal, machine-parsable format on the LLM's input. Providing information in a predictable structure, like a dependency graph, constrains the model's generation, leading to more reliable outputs. This subsection surveys approaches leveraging structural information from the codebase or task to guide the LLM, enhancing its reasoning about complex dependencies.

A prominent strategy uses graph-based representations to capture codebase relationships. Tang \textit{et al.}'s SynFix uses a ``RelationGraph" to map dependencies between code elements~\cite{tang2025synfix}. This constrains the LLM's repair task, ensuring a fix is propagated to dependent components. Similarly, Liu \textit{et al.}'s LLM4TDG defines a ``constraint dependency graph" from testing objectives, converted into prompt constraints to improve test driver generation~\cite{liu2025llm4tdg}.

Another approach imposes a hierarchical structure on generation. Zan \textit{et al.}'s Codes framework generates code repositories from a natural language description (NL2Repo)~\cite{zan2024codes}. It employs a multi-layer sketch, generating a repository sketch (directories, dependencies), then a file sketch (imports, signatures), and finally the implementation, ensuring architectural coherence.

Even for models with long context windows, the input structure remains critical. Peng \textit{et al.} investigated long-context models for repository-level code generation and found that concatenating all relevant files is suboptimal~\cite{peng2025can}. Their study reveals the ordering of code snippets in the prompt significantly impacts performance, suggesting a well-structured prompt reflecting the logical and dependency-based ordering of the code is crucial.

\subsubsection{Interactive}
Interactive prompting extends static prompts into a multi-turn dialogue between the user and the LLM. This conversational approach allows for progressive refinement of solutions. The system can ask clarifying questions to resolve ambiguity, and the user can provide iterative feedback to steer the model toward a correct output. This subsection examines frameworks that facilitate this human-AI collaboration, transforming code generation into a dynamic partnership.

A key motivation for interaction is bridging the semantic gap between user intent and the details needed for code generation. Miao \textit{et al.}'s Clarigen framework identifies ambiguities in a user's request and engages in a clarifying dialogue~\cite{miao2025clarigen}. The question asked to the targeted audience enriches the initial prompt with user feedback prior to generation, improving the alignment with the intent of the user.

Interaction can be user-driven within the IDE. Liu \textit{et al.}'s MACEDON provides real-time code evaluation and optimization suggestions~\cite{liu2025macedon}. A more advanced example is ChatDBG by Levin \textit{et al.}, which augments debuggers with an LLM~\cite{levin2025chatdbg}. This makes the debugger a conversational partner, allowing programmers to ask high-level questions (e.g., ``why is this variable null?") to find a bug's root cause.

However, interaction's effectiveness is not guaranteed. The study by Fu \textit{et al.} employed an iterative framework where new I/O examples could be provided if the code was incorrect~\cite{fu2025first}. They found most successes occurred in the first round, suggesting current LLMs struggle to use iterative feedback when presented as I/O examples. This indicates feedback modality and clarity are critical for successful interaction.
\subsection{Fine-tune-based Solutions}
\label{sec:finetune}

\begin{figure*}[th]
    \centering
    \includegraphics[width=\linewidth]{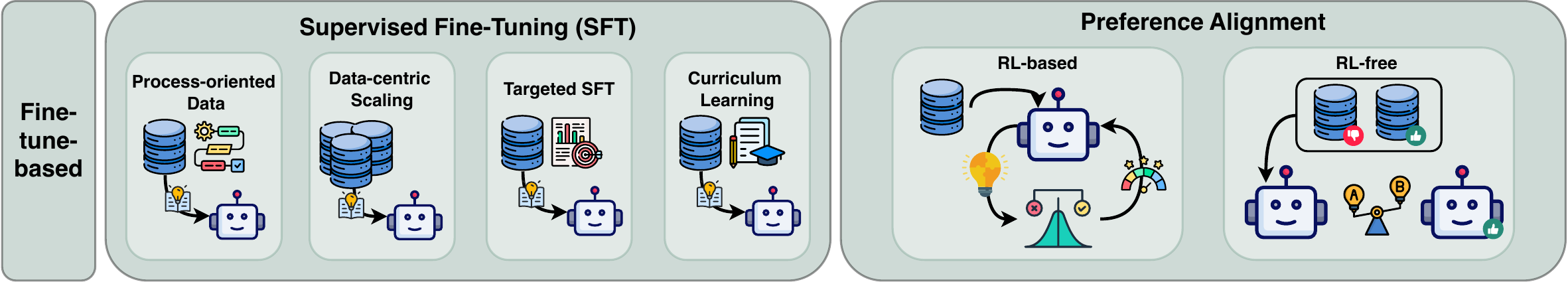}
    \caption{The components of fine-tune-based solutions.}
    \label{fig:finetune}
    \vspace{-0.2in}
\end{figure*}

Fine-tuning is a crucial step beyond prompt engineering, specializing a general-purpose LLM for software engineering. This process involves training a pre-trained model on a curated dataset of domain-specific examples, adapting its parameters to better understand code syntax, semantics, and software development patterns. This section covers primary fine-tuning methodologies, from supervised learning on code-related tasks to preference alignment techniques that teach models to discern high-quality from suboptimal solutions as illustrated in Fig.~\ref{fig:finetune}.

\subsubsection{Supervised Fine-Tuning (SFT)}
Supervised Fine-Tuning (SFT) is the cornerstone of model specialization. It involves training an LLM on a dataset of high-quality input-output pairs, like bug reports and their patches, or natural language descriptions and their code. This process directly teaches the model desired behaviors for specific software engineering tasks. This subsection explores diverse SFT applications, focusing on innovations in dataset creation, curriculum learning strategies, and process-centric training capturing the software development workflow.

A significant SFT trend is shifting from static code snapshots to process-oriented data. Ma \textit{et al.} developed SWE-GPT, fine-tuned on data simulating the software improvement process, including repository understanding, fault localization, and patch generation~\cite{ma2025swe}. Similarly, Piterbarg \textit{et al.} created the D3 dataset, framing code generation as a sequence of file diffs, fine-tuning the model to make iterative changes rather than generating entire files from scratch~\cite{piterbargd3}. Pan \textit{et al.} introduced SWE-Gym, an environment to collect agent's sequences of actions for solving real-world tasks for SFT~\cite{pan2024training}.

Data-centric SFT, focusing on training data quality and scale, is highly effective. Chowdhury \textit{et al.} present SWE-Dev, a methodology creating powerful open-source agents by generating a massive dataset of ``fail-to-pass" test cases from GitHub issues, then used to collect agent trajectories for SFT~\cite{wang2025swe}. This scaling allows smaller models to close the performance gap with larger ones. In formal verification, where data is scarce, Zhang \textit{et al.} developed PoPilot using a synthetic data generation process to teach a model the complex reasoning for proof-oriented programming, enabling a 14B model to outperform GPT-4o~\cite{zhang2025building}. Data curation's importance is highlighted by Cassano \textit{et al.}, who show fine-tuning open Code LLMs on their dataset of code edits with natural language instructions significantly improves their editing capabilities~\cite{cassano2023can}.

Targeted SFT strategies address specific software engineering goals. Yang \textit{et al.} propose MORepair, a multi-objective approach training a model on both the syntactic code transformation (the ``what") and the reasoning behind the fix (the ``why"), for higher-quality repairs~\cite{yang2024morepair}. For bug localization, Chakraborty \textit{et al.} use ``hard example learning" in their BLAZE framework, focusing the SFT process on complex bugs to improve model generalizability~\cite{chakraborty2025blaze}. Sagtani \textit{et al.} improve Fill-in-the-Middle (FIM) code completion by using curriculum learning, creating a dataset of ``hard-to-complete" examples to focus the training where models most often fail~\cite{sagtani2025improving}. For performance optimization, Shypula \textit{et al.} fine-tune a model in their ECO system on a curated dataset of performance anti-patterns and their corresponding optimized solutions mined from historical commits~\cite{shypula2025automated}.

Other SFT approaches include knowledge distillation and leveraging model internals. Chen \textit{et al.} introduce SODA, a framework using a self-paced knowledge distillation cycle to train smaller ``student" models to mimic larger ``teacher" models~\cite{chen2025smaller}. Huang \textit{et al.} developed PtTrust, a risk assessment framework pre-training a risk predictor on a code LLM's internal states to identify incorrect or insecure generations~\cite{huang2025risk}. Architectural innovations also combine with SFT. Yu \textit{et al.} propose the Code Graph Model (CGM), integrating a repository graph representation into the LLM's attention mechanism and using a two-phase SFT process to teach the model structural and semantic dependencies~\cite{tao2025code}. Finally, studies explore SFT's limits. Wang \textit{et al.} found incorporating semantic information from execution traces provided limited SFT benefit, challenging a common assumption~\cite{wang2025code}. Wei \textit{et al.} introduce Co-PatcheR, using an efficient SFT strategy with data distillation and filtering to train small, specialized models for different software patching sub-tasks~\cite{tang2025co}. Project-level context's importance is confirmed by Sapronov \textit{et al.}, who show continued pre-training (a form of SFT) on repository-level data significantly improves code completion~\cite{sapronov2025pretraining}.

\subsubsection{Preference Alignment: RL-based}
Reinforcement Learning (RL)-based alignment is a powerful paradigm for optimizing LLMs against complex, non-differentiable reward signals, like code correctness from passing a test suite. The LLM acts as an agent whose policy (generation strategy) is updated by rewards from an environment or reward model. This allows the model to learn complex reasoning and problem-solving. This subsection examines frameworks using RL, often guided by techniques like Monte Carlo Tree Search, for tasks requiring sophisticated reasoning and planning.

Reward function design is central to RL-based alignment. Several works learn from the reasoning process itself. Zhang \textit{et al.} introduce ReST-MCTS*, a reinforced self-training method using a process-reward model (PRM) to guide an MCTS for generating quality reasoning traces~\cite{zhang2024rest}. The system infers per-step rewards without manual annotation, creating a self-improvement loop fine-tuning the PRM and policy model. Similarly, Zhang \textit{et al.} use SEAlign, using MCTS to evaluate action trajectories and identify ``critical actions" that impact success~\cite{zhang2025sealign}. The model is then fine-tuned on these critical actions with preference optimization, teaching it to favor effective decision-making paths.

Other approaches generate the reward signal dynamically from execution outcomes. Yu \textit{et al.} propose Outcome Refining Process Supervision (ORPS), where the ``reward" is generated at inference time from execution feedback (\textit{e.g.}, correctness, runtime, memory usage) and the LLM's own self-critique~\cite{yu2024reasoning}. The agent explores a tree of thoughts, and this dynamically generated reward signal allows it to learn a preference for more efficient and correct implementation strategies without a pre-trained reward model. This unification of process and outcome rewards represents a significant step towards more autonomous learning.

RL also aligns models with production scenarios. Guo \textit{et al.} use an RL fine-tuning stage to optimize a system for generating functional test scripts~\cite{guo2025optimizing}. After an SFT phase, the model is trained with RL where the reward is based on the similarity between the generated and ground-truth scripts, improving accuracy and reliability. The Kimi K2 model undergoes a joint RL stage, interacting with real and synthetic environments to enhance its agentic capabilities, particularly in tool use~\cite{team2025kimi}.

\subsubsection{Preference Alignment: RL-free}
Preference alignment methods fine-tune models based on judgments of which outputs are better, moving beyond right-or-wrong supervision. RL-free techniques, such as Direct Preference Optimization (DPO), achieve this without reinforcement learning's complexity and instability. They operate on ``chosen" and ``rejected" responses, directly optimizing the model to increase the likelihood of preferred outputs. This subsection reviews methods using RL-free alignment to instill qualities like code security, efficiency, or adherence to coding styles.

A key challenge is sourcing a quality preference dataset. Cassano \textit{et al.} address this with SelfCodeAlign, a self-alignment pipeline that bootstraps preference data~\cite{wei2024selfcodealign}. The base model generates diverse coding tasks and multiple responses for each. It then generates test cases to validate responses in a sandbox. Only instruction-response pairs passing validation are used for the preference dataset, demonstrating a transparent, self-sufficient alignment method.

Innovations are also occurring in the optimization algorithm. Saqib \textit{et al.} introduce Localized Preference Optimization (LPO) to teach LLMs secure coding~\cite{saqib2025teaching}. After distilling a preference dataset of secure vs. insecure code from a teacher model, LPO applies the preference loss only to tokens where the secure and insecure versions diverge. These methods show RL-free alignment is a promising direction for instilling complex attributes into code generation models stably.
\subsection{Agent-based Solutions}
\label{sec:agent}

\begin{figure*}[th]
    \centering
    \includegraphics[width=\linewidth]{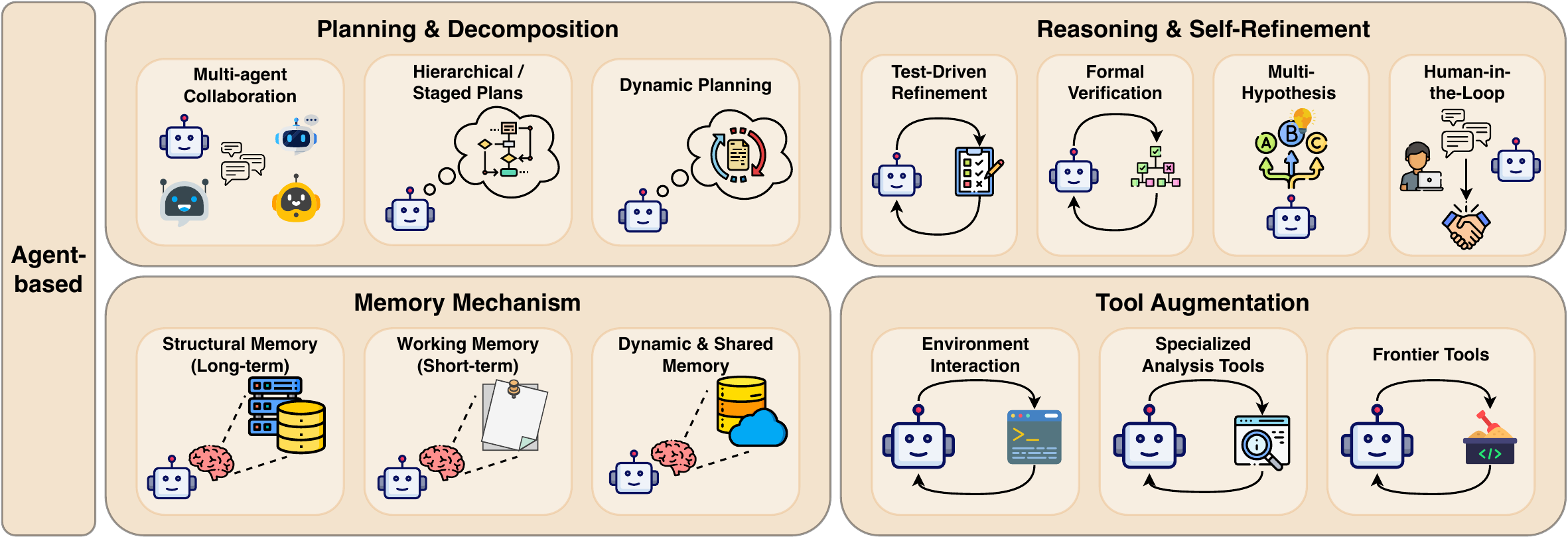}
    \caption{The components of agent-based solutions.}
    \label{fig:agent}
    \vspace{-0.2in}
\end{figure*}

Agent-based systems are the frontier of LLM-driven software engineering, shifting the paradigm from single-shot generation to autonomous, multi-step problem-solving. An ``agent" is a system that can perceive its environment, create and decompose plans, utilize external tools, and learn from feedback to achieve a high-level goal. This approach mimics the iterative and interactive workflow of a human developer, enabling LLMs to tackle complex, repository-level tasks intractable with simple prompting or fine-tuning alone. This section explores the core components of modern software engineering agents, as shown in Fig.~\ref{fig:agent}

\subsubsection{Planning \& Decomposition}
Complex software tasks require breaking down a high-level goal into a sequence of manageable sub-tasks. The Planning and Decomposition component of an agent creates this strategy. Approaches range from static, predefined workflows that guide the agent through fixed stages to dynamic, LLM-driven planning where the agent reasons about its next best action at each step.

A popular approach is multi-agent collaboration, which decomposes tasks by assigning specialized roles. Tao \textit{et al.} introduce MAGIS, a framework with Manager, Developer, and QA Engineer agents mimicking a software team~\cite{tao2024magis}. Nguyen \textit{et al.} model the Agile methodology in AgileCoder, with agents for roles like Product Manager and Tester~\cite{nguyen2025agilecoder}. Wadhwa \textit{et al.} propose MASAI, a modular architecture that decomposes issue resolution into a static pipeline of sub-agents: Test Template Generator, Issue Reproducer, Edit Localizer, Fixer, and Ranker~\cite{arora2024masai}. The Artemis AI framework decomposes code optimization by identifying bottlenecks and distributing these snippets to specialized LLM agents~\cite{giavrimis2025artemis}. Similarly, Xiao \textit{et al.} use a ``divide and conquer" strategy in XpandA, where a long text is partitioned and assigned to multiple ``Explorer" agents~\cite{xiao2025long}. The Co-PatcheR framework uses a collaborative system of smaller, specialized models for localization, generation, and validation~\cite{tang2025co}.

Other approaches use a hierarchical or staged plan. Zan \textit{et al.}'s Codes framework decomposes repository generation into outlining the repository, sketching each file, and then filling the implementation~\cite{zan2024codes}. Wang \textit{et al.} developed LLMDFA, which decomposes dataflow analysis into a three-phase pipeline: source/sink extraction, dataflow summarization, and path feasibility validation~\cite{wang2024llmdfa}. Zhao \textit{et al.} introduce DRCodePilot, which formalizes a ``design first, then code" plan by leveraging design rationales from issue logs before generating the patch~\cite{zhao2024enhancing}. The MarsCode Agent employs a systematic, multi-phase process including planning, reproduction, localization, generation, and validation~\cite{liu2024marscode}. In a vision paper, Shetty \textit{et al.} outline a framework for AIOps agents to handle tasks like root cause analysis through a planned sequence of observing, hypothesizing, and acting~\cite{shetty2024building}.

More dynamic planning mechanisms allow the agent to adapt its strategy. The ExecutionAgent by Bouzenia \textit{et al.} uses a two-phase plan for running tests: a preparation phase to gather information, followed by a feedback loop to refine its plan based on command outputs~\cite{bouzenia2025you}. Wang \textit{et al.} propose ExploraCoder, which decomposes using unseen APIs into subtasks and uses a ``Chain of API Exploration" to rectify its plan based on executability feedback~\cite{wang2024exploracoder}. Ma \textit{et al.} introduce LingmaAgent, which uses Monte Carlo Tree Search (MCTS) to explore a repository knowledge graph, allowing it to plan its information gathering before attempting a fix~\cite{ma2025improving}. This ``understand-then-act" approach is further detailed in RepoUnderstander~\cite{ma2024understand}. The Alphaverus framework decomposes verified code generation into three stages: Exploration, Treefinement, and Critique~\cite{aggarwal2024alphaverus}. In contrast, Xia \textit{et al.} challenge complex planning with ``Agentless," a framework using a fixed, three-phase process (localization, repair, validation), suggesting a simple plan can be sufficient~\cite{xia2025demystifying}. PatchPilot employs a stable, rule-based workflow of reproduction, localization, generation, validation, and refinement~\cite{li2025patchpilot}. Finally, some works focus on observing the planning process itself. Epperson \textit{et al.} developed AGDebugger, a tool for developers to visualize, inspect, and edit messages between agents, allowing modification of the distributed plan~\cite{epperson2025interactive}.

\subsubsection{Reasoning \& Self-Refinement}
Reasoning and self-refinement are the cognitive core of an agent, enabling it to analyze feedback and improve its solutions. This is often implemented as a ``generate-test-revise" loop, where the agent proposes a solution, evaluates it with feedback, and reasons about the outcome to generate a better solution. This subsection explores this feedback loop, from human-in-the-loop systems to autonomous self-critique.

Refinement is most commonly driven by execution and testing feedback. AutoCodeRover by Zhang \textit{et al.} uses an iterative code search and patch generation loop; if a patch fails, the agent retries~\cite{zhang2024autocoderover}. Yu \textit{et al.} designed PATCHAGENT with an integrated verifier that tests patches against security and functional tests; failures trigger a new reasoning cycle~\cite{zheng2025patch}. Chowdhury \textit{et al.} introduce ``iteration scaling" in SWE-Dev, giving the agent more interaction rounds to analyze test failures and refine its code~\cite{wang2025swe}. Hu \textit{et al.} developed Repo2Run, an agent that creates a Dockerfile by iteratively building an image, analyzing errors, and refining it~\cite{hu2025llm}. The MarsCode Agent uses feedback from a sandboxed environment, like exception stacks, to drive its refinement process~\cite{liu2024marscode}. Wang \textit{et al.} found many ``solved" issues in the SWE-bench benchmark are incorrect, calling for more sophisticated validation in agent refinement loops~\cite{wang2025solved}.

A more rigorous form of feedback comes from formal verification. Aggarwal \textit{et al.} created Alphaverus, a self-improving agent using feedback from a formal verifier to correct errors~\cite{aggarwal2024alphaverus}. Its ``Treefinement" mechanism treats refinement as a tree search, allowing it to explore different correction paths.

Human-in-the-loop refinement leverages developer expertise. Takerngsiri \textit{et al.} propose the HULA framework, where an AI generates outputs for a Human Agent to review and correct~\cite{takerngsaksiri2025human}. Omidvar Tehrani \textit{et al.} evaluated a human-AI partnership for code migration, showing human oversight is critical for tasks requiring contextual understanding~\cite{omidvar2024evaluating}. Reiss \textit{et al.} developed ROSE, an IDE-based framework where the developer initiates repair from a debugger, then selects the fix from the agent's suggestions~\cite{reiss2025rose}.

Some agents employ self-critique or explore multiple hypotheses. Zelikman \textit{et al.} introduced Self-Taught Optimizer (STOP), where an ``improver" program recursively optimizes itself by proposing and evaluating modifications to its own code~\cite{zelikman2024self}. Dou \textit{et al.} proposed an iterative self-critique method where an LLM corrects its code based on a bug taxonomy and compiler feedback~\cite{dou2024s}. The Artemis AI framework generates candidate optimizations from multiple LLMs and uses a search and filtering process to select the best solution~\cite{giavrimis2025artemis}. Sohrabizadeh \textit{et al.} use a similar strategy in Nemotron-CORTEXA, generating diverse candidate patches and using self-generated tests and majority voting to select the best one~\cite{sohrabizadehnemotron}. The reasoning in Co-PatcheR is a multi-agent affair where models generate and critique solutions, with a majority vote for the final decision~\cite{tang2025co}.

Other notable reasoning approaches include using simulation for feedback, as demonstrated by Nouri \textit{et al.}, whose agent generates code for autonomous driving systems, which is tested in a simulation environment, with results fed back for refinement~\cite{nouri2025simulation}. Pu \textit{et al.} explore proactive agent reasoning in ``Codellaborator," which analyzes developer context to decide when to offer assistance~\cite{pu2025assistance}. The reasoning process can also be embedded, as in LLM4TDG, which uses constraint reasoning and backtracking to refine generated test drivers~\cite{liu2025llm4tdg}. Finally, the reasoning in CodeV is enhanced by processing visual information, allowing it to use visual feedback to verify a patch, especially for GUI bugs~\cite{zhang2024codev}.

\subsubsection{Memory Mechanism}
To solve repository-level tasks, an agent must overcome LLM context window limitations by managing information. Memory mechanisms provide agents the ability to store, retrieve, and synthesize information from codebases and interaction histories. These mechanisms can be categorized into short-term ``working memory" for context retrieval and long-term ``structural memory" that captures an understanding of the repository. This subsection investigates memory architectures like vector databases, knowledge graphs, and evolving knowledge bases.

Several agents focus on building a repository-scale structural memory for a global understanding of the codebase. Ma \textit{et al.} introduce LingmaAgent, which condenses repository information into a queryable knowledge graph, allowing more informed decisions~\cite{ma2025alibaba}. This concept is further developed in RepoUnderstander, which organizes the repository into a hierarchical tree of files, classes, and functions as the agent's memory~\cite{ma2024understand}. Yu \textit{et al.} take this further with the Code Graph Model (CGM), which integrates a code graph with hierarchical and reference dependencies into the LLM's attention mechanism as a structured memory system~\cite{tao2025code}. Truong \textit{et al.} propose the COLLMS framework, which integrates LLMs with a ``Platform Knowledge" base of software catalogs and code patterns, acting as external memory to compensate for lack of domain-specific information~\cite{truong2024coordinating}.

Other agents rely on on-demand context retrieval to create a ``working memory". Zhang \textit{et al.}'s AutoCodeRover interacts with an AST, using APIs to retrieve code snippets and build contextual memory before patch generation~\cite{zhang2024autocoderover}. Chakraborty \textit{et al.} use dynamic chunking in BLAZE to break source code into indexed segments, allowing the agent to query for code chunks relevant to a bug report~\cite{chakraborty2025blaze}. The AEGIS framework for bug reproduction uses a context construction module that extracts information from the issue description and relevant code, creating a focused memory~\cite{wang2025aegis}. Ehsani \textit{et al.} propose a method to identify knowledge gaps in a prompt, triggering an agent to retrieve information from a knowledge base to fill gaps in its context~\cite{ehsani2025towards}.

The most advanced memory mechanisms are dynamic. Su \textit{et al.} developed EvoR, a retrieval-augmented generation pipeline with a dynamic ``knowledge soup" memory. This memory is updated with information from web searches, documentation, and execution feedback, allowing the agent to learn and adapt~\cite{su2024evor}. For multi-agent systems, memory must be coordinated. Xiao \textit{et al.} use a centralized shared memory in their XpandA framework, updated via a ``question-guided protocol" to maintain consistent understanding across collaborating agents~\cite{xiao2025long}.

\subsubsection{Tool Augmentation}
Tool augmentation grounds an agent's capabilities by giving it the ability to interact with the software development environment. By equipping agents with tools—such as terminal access, file editors, and compilers—they can perform actions and gather information impossible with text generation alone. This subsection overviews tool-use strategies, from Agent-Computer Interfaces to integrating static analysis, formal verification, and multimodal inputs.

A fundamental tool use is interacting with a development environment like a terminal or IDE. Yang \textit{et al.} introduced SWE-agent, centered on an Agent-Computer Interface (ACI) with commands tailored for software engineering, like a file viewer and editor, proving more effective than general terminal access~\cite{yang2024swe}. Bouzenia \textit{et al.}'s ExecutionAgent uses terminal commands to explore and run build systems, and ``meta-prompting," where an LLM generates guidelines for using specific technologies~\cite{bouzenia2025you}. Hu \textit{et al.}'s Repo2Run agent depends on the Docker engine and testing frameworks; its workflow is a loop of sending commands and parsing output to refine a Dockerfile~\cite{hu2025llm}. The MarsCode Agent utilizes code knowledge graphs and the Language Server Protocol to navigate and analyze code like a developer using an IDE~\cite{liu2024marscode}. Li \textit{et al.} propose mediator agents in IDEs to enhance human-developer interaction with LLMs~\cite{li2025enhancing}.

Many agents are augmented with specialized analysis tools. Zhang \textit{et al.}'s AutoCodeRover integrates Spectrum-Based Fault Localization (SBFL) to analyze test outcomes and assign suspicion scores, directing the bug search~\cite{zhang2024autocoderover}. Wang \textit{et al.}'s LLMDFA uses the LLM as a ``tool-builder," generating scripts that use parsing libraries to identify sources/sinks and the Z3 theorem prover to validate dataflow paths~\cite{wang2024llmdfa}. Nguyen \textit{et al.}'s AgileCoder uses a ``Dynamic Code Graph Generator" to provide agents an evolving representation of the software's structure~\cite{nguyen2025agilecoder}. Kitsios \textit{et al.} developed BLAST, a tool that combines LLM with Search-Based Software Testing (SBST), where the LLM generates a ``seed" test that the SBST tool evolves~\cite{kitsios2025automated}.

The frontier of tool use involves formal verification, simulation, and multimodal inputs. Aggarwal \textit{et al.}'s Alphaverus depends on an external formal verifier; its error messages are the primary tool governing the agent's action loop and ``Treefinement" search~\cite{aggarwal2024alphaverus}. PatchPilot also uses formal verification tools for higher assurance~\cite{li2025patchpilot}. Nouri \textit{et al.} integrate their agent with a simulation model for autonomous driving to evaluate generated code in realistic scenarios~\cite{nouri2025simulation}. Expanding sensory input, Zhang \textit{et al.} created CodeV, an agent that processes screenshots and videos from issue reports to resolve GUI bugs difficult to diagnose from text~\cite{zhang2024codev}. Soni \textit{et al.} introduced OpenHands-Versa, a generalist agent that uses a minimal set of tools, including multimodal web browsing and file access~\cite{soni2025coding}.

Finally, the design of tools and the environment is a research area. Pan \textit{et al.}'s SWE-Gym provides an interactive environment with thousands of Python tasks for training and evaluating tool-using agents~\cite{pan2024training}. The Kimi K2 model is optimized for agentic tool use, capable of orchestrating commands to render, test, and debug code~\cite{team2025kimi}.
\section{Benchmarks for Real-World Software Engineering}
\label{sec:benchmarks}
Benchmarks have long been a core tool for advancing LLM research for software engineering. From early function-level code generation datasets, such as HumanEval~\cite{chen2021evaluating} and MBPP~\cite{austin2021program}, to large-scale multilingual benchmarks, such as CodeNet~\cite{puri2021codenet} , these efforts have significantly advanced LLM development. However, most of these efforts focus on fragmented tasks, which only partially capture the complexity of real-world software engineering. In practice, software systems involves complex processes such as repository-level generation, bug reporting, testing, and integration. This gap has motivated the creation of real-world code benchmarks, such as SWE-bench~\cite{jimenez2024swebench} and its successors, which incorporate real GitHub issues into their testing environments, allowing them to evaluate the performance of models and agents in scenarios more closely resembling those of real software development. This chapter reviews these benchmarks based on their focused tasks, including code generation, code translation, program repair, and other tasks.

\begin{table}[t]
\centering
\caption{Representative Benchmarks for three SE tasks (1. Code Generation, 2. Code Translation, and 3. Program Repair) at Function- and Repository-Level. $^{*}$SWE-Bench reports 2,294 instances; each instance may generate multiple test cases.}
\label{tab:benchmark}
\begin{tabular}{lcccc}
\toprule
Benchmark  & Tasks & \#PLs                     & Scope      & \#Tests \\
\midrule
HumanEval  & 1     & 1 (Python)                & Function   &  164    \\
DevEval    & 1     & 1 (Python)                & Repository &   1,874   \\
CodeXGLUE  & 1,2,3 & \makecell{10 (Python, Java, \\ C++, C, etc.)}         & Function   & $>$ 100K     \\
AlphaTrans & 2     & 2 (Java, Python)          & Repository & 2,719     \\
SWE-Bench  & 3     & 1 (Python)                & Repository &  2,294$^{*}$    \\
\bottomrule
\end{tabular}
\end{table}

\subsection{Code Generation}
\label{sec:code_generation}
Code generation refers to automatically generating program code according to natural language descriptions. 
Pioneering benchmarks in code generation, such as HumanEval~\cite{chen2021evaluating}, MBPP~\cite{austin2021program}, and CodeXGLUE~\cite{lu2021codexglue}, typically focused on function-level tasks, evaluating the ability of language models to generate syntactically correct and functionally equivalent code implementations. To support testing on more programming languages, Puri \textit{et al.}~\cite{puri2021codenet} proposed CodeNet covering 55 programming languages.
However, these benchmarks often rely on relatively simple tasks and limited testings, which fail to reflect real-world software engineering situations. To mitigate this issue, Liu \textit{et al.}~\cite{Liu2023is} proposed HumanEval+, extending HumanEval~\cite{chen2021evaluating} by 80x with synthetic test cases.

Subsequent work has enriched evaluation along multiple dimensions. Chai \textit{et al.}~\cite{chai2025mceval} proposed McEval, covering 40 PLs with 16,000 samples and supporting tasks beyond code generation, including code review, and error detection. Similarly, Yan \textit{et al.}~\cite{yan2024codescope} proposed CodeScope, covering 43 programming languages and 8 coding tasks.
While, Yu \textit{et al.}~\cite{yu2025humanevalpro} extended the HumanEval and MBPP benchmarks by introducing a self-invoking code generation task. This task requires models to first solve a base problem and then utilize its solution to address a related but more complex problem.
Meanwhile, Zhuo \textit{et al.}~\cite{zhuo2025bigcodebench} developed BigCodeBench, focusing on evaluating LLM on handling complex instructions and diverse function call. 
In terms of evaluation scope, Du \textit{et al.}~\cite{du2024classeval} introduced class-level generation tasks in ClassEval.
Gong \textit{et al.}~\cite{gong2024evaluation} constructed a benchmark for the Fill-in-the-Middle (FIM) completion.                 
Beyond functional correctness, Qing \textit{et al.}~\cite{qing2025effibench} proposed EffiBench-X, highlighting efficiency gaps between LLMs and human experts
Thakur \textit{et al.}~\cite{thakur2025clever} introduced formal verifiability in CLEVER, which requires generated code to pass Lean's type checker.     

To better align with realistic software development processes, 
Inspired by SWE-Bench, Vergopoulos \textit{et al.}~\cite{vergopoulos2025automated} proposed an automated system SetUpAgent that can construct repository-level code benchmarks from GitHub repositories and released an extended SWE-bench. Li \textit{et al.}~\cite{li2024deveval} proposed DevEval including 117 repositories with munual annotations, aiming to preserve realistic code distributions and dependency distributions. Similarly, Liang \textit{et al.}~\cite{liang2025repocod} created RepoCOD sourced from 11 real-world projects. Recognizing the evolving nature of software systems, Zheng \textit{et al.}~\cite{Zheng2025humanevo} developed an evolving repository-level code generation benchmark, named HumanEvo. It evaluates the LLM performance across project evolution by treating multiple project versions as contextual sources. In addition, Jain \textit{et al.}~\cite{jain2024r2e} proposed a framework that can transform GitHub repositories into interactive environment and released R2E-Eval benchmark to evaluate LLM-based agents on code generation. Recently, Yang \textit{et al.}~\cite{yang2025swesmith} and Badertdinov \textit{et al.}~\cite{badertdinov2025swe} proposed SWE-smith and SWE-rebench, respectively, both significantly extending SWE-Bench~\cite{jimenez2024swebench} to benchmark LLM agents under more diverse and realistic settings.

Beyond functionality-oriented evaluation, Chen \textit{et al.}~\cite{chen2025dycodeeval} focused on data contamination issues and proposed DyCodeEval, which benchmarks LLMs under varying contamination levels. 
Considering the ambiguity, inconsistency, and incompleteness often present in real-world software engineering problem descriptions, Wu \textit{et al.}~\cite{wu2025humanevalcomm} designed HumanEvalComm to evaluate the communication skills of LLMs in code generation tasks. Similarly, Fu \textit{et al.}~\cite{fu2025first} proposed InterCode, which evaluates code generation using only I/O examples, thereby emphasizing the need for LLMs to understand and implement incomplete and diverse code requirements. In addition, Wang \textit{et al.}~\cite{wang2025maintaincoder} proposed MaintainCoder, which emphasizes that generated code should not only be functionally correct but also meet maintainability standards.

\subsection{Code Translation}
\label{sec:code_gtranslation}
Code translation, which aims to covert code from one programming language (PL) to another while preserving its functionality, is a critical task in modern software engineering. It enables the migration of legacy systems, facilitates interoperability across heterogeneous platforms, and supports cross-language software development. Leveraging their ability to capture complex code semantics and contextual dependencies, LLms have emerged as powerful tools for advancing the automation of code translation.

To systematically evaluate the capability of LLMs in this task, various benchmarks have been established. CodeTrans in CodeXGLUE ~\cite{lu2021codexglue} provided parallel Java–C\# code from open-source projects. CodeNet~\cite{puri2021codenet} further extended coverage to 55 PLs. Ahmad \textit{et al.}~\cite{ahmad2023avatar} presented AVATAR for Java–Python translation, providing unit tests for 250 evaluation examples to measure functional accuracy. In general, these works mainly focusing on function-level tasks.

Building on these datasets, subsequent benchmarks expanded both language coverage and evaluation methodology. Yan \textit{et al.}~\cite{yan2023codetransocean} constructed CodeTransOcean supporting code translation across 8 PLs and including a dataset for translating deep learning code across different frameworks. They also introduced a execution-based evaluation metric \texttt{Debugging Success Rate @K} to evaluate the performance of LLMs. Similarly, Zheng \textit{et al.}~\cite{zheng2023codegeex} and Khan \textit{et al.}~\cite{khan2024xcodeeval} released HumnEval-X and xCodeEval, respectively, and adopted execution-based evaluation to address the shortcomings of similarity metrics such as BLEU~\cite{papineni2002bleu} and CodeBLEU~\cite{lu2021codexglue}. Scaling up, Nie \textit{et al.}~\cite{nie2023crosscodebench} provided 216 tasks across 18 PLs in CodeCrossBench to assess generalization across diverse software engineering tasks. While Tao \textit{et al.}~\cite{tao2024unraveling} extended HumanEval~\cite{chen2021evaluating} into PolyHumanEval, covering 14 PLs for large-scale multilingual evaluation.

Driven by real-world migration needs, benchmarks then advanced to class-level and repository-level translation.
Xue \textit{et al.}~\cite{du2024evaluating} extended ClassEval into ClassEval-T, supporting Python to Java and C++ class-level tasks. At the repository level, several large-scale benchmarks have emerged, such as CRUST-Bench~\cite{khatry2025crustbench}, RustRepoTrans~\cite{ou2024repository}, and RepoTransBench~\cite{wang2024repotransbench}, each introduced extensive translation tasks from real-world GitHub repositories. Further advancing this direction, Ibrahimzada \textit{et al.}~\cite{ibrahimzada2025alphatrans} proposed AlphaTrans, a neuro-symbolic framework to automate repository-level code translation and applied it to translate ten real-world open-source projects.

Beyond functional correctness, research has also expanded to assess translation quality and robustness. Zhang \textit{et al.}~\cite{zhang2025functiontostyle} proposed F2STRANS, which includes manual annotations to support stylistic readability evaluation. While Guo \textit{et al.}~\cite{guo2025codeeditorbench} presented CodeEditorBench to evaluate broader editing and transformation capabilities.
Jiao \textit{et al.}~\cite{jiao2023on} established a four-level taxonomy of translation tasks, including token, syntactic, library, and algorithm levels, and built G-TransEval for integrated evaluation.

\subsection{Program Repair}
\label{sec:program_repair}
Automated Program Repair(APR) aims to software reliability by automatically locating and fixing software bugs. The development of LLMs has transformed the APR research paradigm and evaluation methodologies, promoting the development of diverse benchmarks that assess the capabilities of APR technologies at different levels and in different scenarios.   

To test LLM debugging capabilities, Tian \textit{et al.}~\cite{tian2024debugbench} collected 4253 instances from LeetCode across C++, Java, and Python and constructed DebugBench, covering syntax, reference, logic, and multiple error cases. As focus shifted to more complex settings, Jimenez \textit{et al.}~\cite{jimenez2024swebench} introduced SWE-Bench, which is directly derived from GitHub issues and their corresponding pull requests. It requires LLMs to generate patches to modify the codebase, with correctness verified against real test cases. 
Chen \textit{et al.}~\cite{chen2024when} and Liu \textit{et al.}~\cite{liu2025repodebug} introduced RepoBugs and RepoDebug, with realistic repository-level errors and multi-task debugging data, highlighting the importance of cross-file dependencies and context extraction methods. 
Lee \textit{et al.}~\cite{Lee2024the} proposed GitHub Recent Bugs (GHRB), which collects real-world Java bugs from most stars Java repositories in GitHub. 
Rando \textit{et al.}~\cite{rando2025longcodebench} introduced LongCodeBench, targeting code comprehension and repair with a context window of up to 1M tokens, revealing the bottlenecks of long-context models in real-world scenarios. Similarly, Hariharan \textit{et al.}~\cite{hariharan2025breakpoint} injected adversarial errors into real-world GitHub repositories, constructing multi-dimensional tasks ranging from local to system-level reasoning to stress-test the repair capabilities of LLM agents. These benchmarks highlight the necessity of agentic approaches that integrate planning, decomposition, and tool use.

As the software ecosystem diversifies, benchmarks are also increasingly targeting multilingual and multimodal tasks. Zan \textit{et al.}~\cite{zan2025multiswebench} extended SWE-bench~\cite{jimenez2024swebench} to 8 PLs, including Python, Java, TypeScript, JavaScript, Go, Rust, C and C++, enabling cross-ecosystem comparisons. Their study revealed that current LLMs perform well on Python-related issues but demonstrate limited generalization capability across other PLs. Guo \textit{et al.}~\cite{guo2025omnigirl} introduced a benchmark OmniGIRL including not only textual but also multimodal information such as images in the issue descriptions, providing a multilingual, multimodal, and multi-domain evaluation setting.

In real-world deployment scenarios, benchmarks increasingly emphasize security. Wang \textit{et al.}~\cite{wang2025cve} proposed an evaluation framework CVE-Bench collecting 509 common vulnerabilities and exposures across four PLs to evaluate LLM-based agents' abilities in a realistic vulnerability-repairing scenarios. Targeting security engineering, Lee \textit{et al.}~\cite{lee2025sec} developed SEC-Bench for proof-of-concept (PoC) generation and vulnerability patching tasks. They found that the current models can only achieve a success rate of at most 18\% and 34\% in these two tasks, illustrating the complexity of security-related repair. While Nie \textit{et al.}~\cite{yang2024seccodeplt} provided over 5.9k samples across 44 CWE-based risk categories in SECCODEPLT, supporting evaluations of secure code generation, vulnerability detection, and patching.

\subsection{Other benchmarks}
\label{sec:others}
This section reviews benchmark studies that may not directly target code output, but focus on aspects that are also important for real-world software development, such as code understanding, dependency installation, and collaborative development.

Beyond code generation, code understanding and reasoning are also important tasks in software development. Gu \textit{et al.}~\cite{gu2024cruxeval} and Xu \textit{et al.}~\cite{xu2025cruxeval} proposed CRUXEval and CRUXEval-X, respectively, assessing the reasoning capability via input/output prediction tasks. While, Zhao \textit{et al.}~\cite{zhao2025codejudge} introduced CodeJudge-Eval, which evaluates whether LLMs can assess code to evaluate their reasoning capability.
To address the gap in long-code understanding, Li \textit{et al.}~\cite{li2025longcodeu} proposed LONGCODEU, which tests four aspects of comprehension, including code unit perception, intra-code unit understanding, intercode unit relation understanding, and long code documentation understanding. 
Recently, He \textit{et al.}~\cite{he2025tf} introduced TF-Bench, which focused on program semantics reasoning tasks leveraging type inference in a formal natural deduction system, System \textit{F}. 
Xie \textit{et al.}~\cite{xie2025core} with over 12k verified instances, targets static analysis capabilities. 


Targeting test generation, Mundler \textit{et al.}~\cite{mundler2024swtbench} developed SWT-bench, which collects popular Python repositories requiring test generation aligned with user issues. Wang \textit{et al.}~\cite{wang2025testeval} proposed TestEval, with a focus on generating test cases that cover specific program lines, branches, or paths. Building upon SWE-Bench~\cite{jimenez2024swebench}, Jain \textit{et al.}~\cite{jain2025testgeneval} developed TestGenEval, a large-scale dataset containing over 60,000 test cases from GitHub repositories. 

Targeting code refactoring, Gautam \textit{et al.}~\cite{gautam2025refactorbench} built RefactorBench, a benchmark with multi-file refactoring tasks that requires LLMs to manage cross-files dependencies and follow natural language instructions.
Shetty \textit{et al.}~\cite{shetty2025gso} introduced optimization tasks into evaluation by constructing GSO, a benchmark comprising 102 instances from 10 codebases. The benchmark requires models to generate patches improving runtime efficiency of while maintaining functional correctness, with results compared against expert developer optimizations.
For dependency analysis, Zhang \textit{et al.}~\cite{zhang2025di} introduced DI-Bench to evaluate dependency inference ability of LLMs, while Milliken \textit{et al.}~\cite{milliken2025beyond} proposed Installamatic to evaluate repository-level environment setup. 
Addressing real-world collaborative scenarios, Guo \textit{et al.}~\cite{guo2025syncmind} created SynBench to evaluate ability of agents to resolve collaboration conflicts and maintain consistency in asynchronous multi-contributor environments.

Overall, benchmarks for software engineering have evolved from function-level evaluations to repository-level and real-world settings, progressively emphasizing robustness, multilinguality, and practical usability. Table~\ref{tab:benchmark} provides a comparison of representative datasets for three SE tasks at function- and repository-level.
\section{Applications}

\label{sec:applications}

Artificial intelligence, particularly the advent of Large Language Models (LLMs), has catalyzed a paradigm shift across the software engineering landscape, introducing powerful new capabilities for automation and assistance. These AI-driven applications are not confined to a single activity but span the entire software development lifecycle (SDLC), fundamentally reshaping how developers design, write, debug, and maintain code. To systematically present these diverse applications, this section is structured along the sequential phases of the SDLC, illustrating the application of artificial intelligence from the initial stages of requirements analysis and architectural design, through code generation and quality improvement, and extending to testing, verification, and security. The taxonomy of applications is detailed in Table \ref{tab:taxonomy-application}.

\begin{table*}[htbp]
    \centering
    \caption{Taxonomy of Applications for Software Engineering}
    \renewcommand{\arraystretch}{1.2}
    \setlength{\tabcolsep}{5pt}
    \begin{tabular}{>{\raggedright\arraybackslash}p{0.10\textwidth} p{0.14\textwidth} p{0.24\textwidth} p{0.44\textwidth}}
        \toprule
        \textbf{Category} & \textbf{Target} & \textbf{Methods} & \textbf{Related Works} \\
        \midrule
        \textbf{{Analysis \&} \quad Design} &
        Refine ambiguous requirements and formulate high-level architectural strategies &
        Transform abstract inputs into functional code, utilize interactive agents to clarify user instructions, and generate high-level plans before coding &
        PACGBI~\cite{sarschar2024pacgbi}{, }~\cite{kc2025demystifying}{, }DRCodePilot~\cite{zhao2024enhancing}{, }Clarigen~\cite{miao2025clarigen}{, }AgileCoder~\cite{nguyen2025agilecoder}{, }~\cite{pandini2025exploratory} {, }HULA~\cite{takerngsaksiri2025human}{, }Codes~\cite{zan2024codes}{, }PatcheR~\cite{tang2025co} \\ \hline
        \textbf{Code Generation \& \quad Transformation} &
        Automate the creation and modification of source code &
        Enhance IDE tools with context-aware suggestions, generate complete code blocks from instructions, and automate large-scale code migrations and modernizations &
        TGen~\cite{mathews2024test}{, }~\cite{peng2025can}{, }Codes~\cite{zan2024codes}{, }~\cite{sapronov2025pretraining}{, }   COLLMS~\cite{truong2024coordinating}{, }MuSL~\cite{ke2025mutual}{, }Codellaborator~\cite{pu2025assistance}{, }ExploraCoder~\cite{wang2024exploracoder}{, }~\cite{omidvar2024evaluating}   ReLoc~\cite{lyu2025let}{, }AutoCodeRover~\cite{zhang2024autocoderover} {, }   OpenHands-Versa~\cite{soni2025coding}{, }MAGIS~\cite{tao2024magis}{, }   AgileCoder~\cite{nguyen2025agilecoder}{, }EG-CFG~\cite{lavon2025execution}{, }   Acereason~\cite{chen2025acereason}{, }ORPS~\cite{yu2024reasoning}{, }   HULA~\cite{takerngsaksiri2025human}{, }MaintainCoder~\cite{wang2025maintaincoder}{, }   Alphaverus~\cite{aggarwal2024alphaverus}{, }~\cite{nouri2025simulation}{, }~\cite{sagtani2025improving} \\  \hline
        \textbf{Bug Repair \& Debugging} &
        Pinpoint, reproduce, and automatically fix bugs in the codebase &
        Localize faults based on bug reports, automatically generate test cases to reproduce errors, and employ autonomous agents for bug resolution &
        BLAZE~\cite{chakraborty2025blaze}{, }AEGIS~\cite{wang2025aegis}{, }CodeV~\cite{zhang2024codev}{, }SWE-agent~\cite{yang2024swe}{, }PATCHAGENT~\cite{zheng2025patch}{, }SynFix~\cite{tang2025synfix}{, }ChatDBG~\cite{levin2025chatdbg}{, }CORTEXA~\cite{sohrabizadehnemotron}{, }AEGIS~\cite{wang2025aegis}{, }LingmaAgent~\cite{ma2025improving}{, }MORepair~\cite{yang2024morepair}{, }AutoCodeRover~\cite{zhang2024autocoderover}{, }MAGIS~\cite{tao2024magis}{, }ROSE~\cite{reiss2025rose} \\  \hline
        \textbf{Refactoring \& Improvement} &
        Proactively enhance code quality and optimize performance &
        Automatically rewrite code for efficiency, improve code maintainability and readability through dataflow analysis, and address technical debt &
        Artemis~\cite{giavrimis2025artemis}{, }~\cite{shypula2025automated}{, }Afterburner~\cite{du2025afterburner}{, }LLMDFA~\cite{wang2024llmdfa}{, }MaintainCoder~\cite{wang2025maintaincoder}{, }MACEDON~\cite{liu2025macedon}{, }SWE-GPT~\cite{ma2025swe}{, }MuSL~\cite{ke2025mutual}{, }PATCHAGENT~\cite{zheng2025patch} \\  \hline
        \textbf{Testing \& Security} &
        Ensure code correctness, robustness, and security against vulnerabilities &
        Automate test generation and execution, detect and enhance code security through vulnerability analysis, and utilize formal verification for provably correct code in critical systems &
        ExecutionAgent~\cite{bouzenia2025you}{, }CBR~\cite{guo2025optimizing}{, }LLM4TDG~\cite{liu2025llm4tdg}{, }~\cite{nouri2025simulation}{, }Wedge~\cite{yang2025synthesizing}{, }BLAST~\cite{kitsios2025automated}{, }SAGA~\cite{ma2025rethinking}{, }PtTrust~\cite{huang2025risk}{, }LPO~\cite{saqib2025teaching}{, }Alphaverus~\cite{aggarwal2024alphaverus}{, }PoPilot~\cite{zhang2025building}{, }LLMDFA~\cite{wang2024llmdfa}{, }CGM~\cite{tao2025code}{, }REAL~\cite{yao2025training}{, }PurpCode~\cite{liu2025purpcode}{, }PatchPilot~\cite{li2025patchpilot}  \\
        \bottomrule
    \end{tabular}
    \label{tab:taxonomy-application}
\end{table*}

\subsection{Requirements Analysis \& Software Design}
At the inception of the software lifecycle, AI models are being employed to bridge the gap between human intent and technical specification, refining ambiguous requirements and shaping high-level architectural strategy. In the domain of \textit{Requirement Understanding \& Refinement}, research focuses on transforming abstract inputs, such as user stories or feature requests from a product backlog, directly into functional code, thereby minimizing manual translation~\cite{sarschar2024pacgbi}. To achieve this, some systems can automatically ``purify" and structure vague user requests into complete, actionable descriptions for developers~\cite{kc2025demystifying}. Other approaches create interactive agents that can proactively ask clarifying questions to resolve ambiguities in user instructions before proceeding with implementation~\cite{miao2025clarigen}. This capability can be embedded within larger multi-agent frameworks that simulate agile development roles, such as a product owner, to autonomously handle the entire cycle from requirement grooming to final delivery~\cite{nguyen2025agilecoder}.

Following requirements gathering, AI is also being applied to \textit{High-level Solution \& Architectural Design}. A key insight in this area is that generating a high-level plan or design before writing code significantly improves outcomes. This ``solution design" phase allows models to first outline a repair strategy in natural language or pseudo-code, mirroring human expert behavior~\cite{zhao2024enhancing}. This planning stage can be interactive, allowing human engineers to guide and correct an agent's high-level plan before code is generated~\cite{takerngsaksiri2025human}. For new projects, models can first generate the entire repository structure and file skeletons as a ``multi-layer sketch" before populating the implementation details~\cite{zan2024codes}. Furthermore, AI can analyze existing codebases to identify architectural ``smells"—such as cyclic dependencies or excessive coupling—and propose appropriate refactoring strategies~\cite{pandini2025exploratory}. This capability could be integrated into CI/CD pipelines as an ``architecture guardian" to prevent design flaws or used by a ``commander" agent to coordinate complex, multi-component bug fixes~\cite{pandini2025exploratory, tang2025co}.

\subsection{Code Generation \& Transformation}
Once requirements and designs are established, AI's role transitions to the automated generation and transformation of source code, one of its most mature application areas. In \textit{Code Completion \& Assisted Writing}, models enhance IDE tools with more contextually aware suggestions. This includes ``fill-in-the-middle" (FIM) capabilities, where the model predicts missing code segments within an existing block~\cite{sagtani2025improving}. To improve accuracy, research has focused on pretraining models for project-level completion by using code graph structures to better understand repository-wide context~\cite{sapronov2025pretraining}. This enables proactive assistants that not only complete code but also suggest refactorings or warn of potential errors ~\cite{pu2025assistance}.

Beyond mere completion, AI excels at \textit{Functional Code Synthesis}, where complete blocks of code or even entire applications are generated from instructions. A prominent paradigm is test-driven development (TDD), where models generate functional code that satisfies a set of pre-written unit tests~\cite{mathews2024test}. The scope of generation has expanded significantly, with long-context models capable of implementing complex, cross-cutting features at the repository level~\cite{peng2025can}, and some systems aiming to generate a complete, functional code repository from a single natural language sentence~\cite{zan2024codes}. This is often accomplished by autonomous agents that can handle end-to-end software evolution tasks, sometimes organized into multi-agent frameworks that simulate agile development methodologies to tackle complex requirements~\cite{zhang2024autocoderover, tao2024magis, nguyen2025agilecoder}. To handle novel problems, these agents can learn to use unseen APIs by exploring documentation~\cite{wang2024exploracoder} or browse the web to find solutions for configuration errors and library usage~\cite{soni2025coding}. The generation process itself is improved through various techniques, such as iterative self-correction based on execution feedback~\cite{lyu2025let, lavon2025execution}, optimizing reasoning with reinforcement learning~\cite{chen2025acereason, yu2024reasoning}, and interacting with humans to clarify instructions~\cite{takerngsaksiri2025human, miao2025clarigen}. For specialized domains, research is focused on generating code with specific properties, such as high maintainability to accommodate dynamic requirements~\cite{wang2025maintaincoder}, or generating provably correct code for safety-critical systems through simulation-guided refinement or by using formal proofs as an intermediate step~\cite{nouri2025simulation, aggarwal2024alphaverus}.

Another function is \textit{Code Transformation \& Migration}, where AI automates the often tedious process of updating or translating codebases. This includes large-scale migrations, where human-AI partnerships can handle the bulk of repetitive work, such as upgrading a project to a new framework version~\cite{omidvar2024evaluating}. This is also applied to software modernization, such as migrating a legacy system to a new cloud platform by coordinating LLMs with platform-specific knowledge~\cite{truong2024coordinating}. More granularly, AI tools can perform sound static analysis to safely migrate code from deprecated I/O APIs to modern equivalents or automatically parallelize sequential C/C++ code into high-performance CUDA code to leverage modern hardware~\cite{ke2025mutual}. The potential for this technology extends to automating library upgrades and other code-base modernizations~\cite{zhang2024autocoderover}.

\subsection{Bug Repair \& Debugging}

Beyond initial code creation, AI offers significant support in the critical phase of bug repair and debugging. A crucial first step is \textit{Bug Localization \& Reproduction}. Models can pinpoint the exact location of a fault within a large codebase based on a natural language bug report, even across different languages and projects~\cite{chakraborty2025blaze}. This localization can be enhanced by providing agents with better repository exploration capabilities or by improving their ability to identify the precise ``pain points" that need modification~\cite{sohrabizadehnemotron}. Once a location is identified, other systems automatically reproduce the bug by generating an executable test case that triggers the failure~\cite{wang2025aegis}. This process can be enhanced by incorporating multimodal data; for instance, models can leverage screenshots from a bug report to better understand and reproduce UI-related issues~\cite{zhang2024codev}.

With a bug localized, the focus shifts to \textit{Automated Program Repair} (APR). Here, autonomous agents are designed to resolve software engineering tasks end-to-end, fixing real-world issues in GitHub repositories~\cite{yang2024swe}. These agents often mimic human workflows by analyzing the repository for context, examining the code's dependency graph to ensure patch correctness, and even considering multiple objectives like correctness and code quality~\cite{ma2025improving, tang2025synfix, yang2024morepair}. Advanced agents can handle the full lifecycle of a fix, from creating a new code branch to generating the appropriate patch and writing a commit message, demonstrating the potential for fully autonomous software maintenance~\cite{zhang2024autocoderover, tao2024magis}.

Complementing fully automated repair, AI is also enhancing \textit{Interactive Debugging Assistance}. Tools like ChatDBG augment traditional debuggers with a conversational interface, allowing a developer to query the program state in natural language (e.g., ``why is x null?") after a crash~\cite{levin2025chatdbg}. Similarly, IDE-integrated frameworks like ROSE bring the developer ``into the loop," allowing them to interact with the repair tool to provide feedback and guide the generation of solutions, making the debugging process more intuitive and efficient~\cite{reiss2025rose}.

\subsection{Code Refactoring \& Quality Improvement}
Maintaining software quality extends beyond fixing bugs, leading to AI applications in proactive code refactoring and optimization. A key area of focus is \textit{Code Performance Optimization}, where AI systems automatically rewrite code to be more efficient. This can be achieved through multi-LLM frameworks that collaborate to improve performance~\cite{giavrimis2025artemis} or by systems that specialize in optimizing high-level scripts for data warehouses~\cite{shypula2025automated}. More advanced frameworks utilize reinforcement learning, allowing a model to continuously improve its optimization ability by interacting with the execution environment and learning from the performance outcomes of its changes~\cite{du2025afterburner}. This also includes translating sequential code to high-performance parallel code, a direct method for performance enhancement~\cite{ke2025mutual}.

Alongside performance, AI is being used for \textit{Improving Code Maintainability \& Readability}. Models can perform complex dataflow analysis to help developers understand code by answering natural language queries like ``Where does this variable's value come from?"\cite{wang2024llmdfa}. Other approaches focus on generating code that is inherently more maintainable and adaptable to future changes in requirements\cite{wang2025maintaincoder}. This capability can be delivered directly within the IDE, where plugins provide real-time, multi-dimensional feedback on code quality—assessing readability, complexity, and performance—and proactively suggest improvements~\cite{liu2025macedon}. Broader, process-centric models aim for holistic software improvement, capable of handling tasks from refactoring to implementing new features~\cite{ma2025swe}. This includes agents designed to address technical debt by identifying and fixing ``bad smells" accumulated in the codebase~\cite{zheng2025patch}.

\subsection{Testing, Verification, \& Security}
To ensure correctness and robustness, AI is increasingly applied to software testing, verification, and security. In \textit{Automated Test Generation \& Execution}, agents are being developed to execute tests for arbitrary projects without needing manual configuration, simplifying CI/CD setup~\cite{bouzenia2025you}. These tools can automatically generate functional test scripts, create complex test data that satisfies specific logical constraints, and synthesize performance test cases to evaluate code efficiency~\cite{guo2025optimizing, liu2025llm4tdg, yang2025synthesizing}. A particularly powerful application is the automated generation of tests that reproduce a specific issue from a bug report, which is invaluable for regression testing~\cite{kitsios2025automated}. Some frameworks even unify code and test generation, pushing towards a model of self-verification where the LLM is also responsible for generating challenging tests for its own code~\cite{ma2025rethinking}.

For \textit{Code Security Enhancement \& Vulnerability Detection}, research is exploring multiple defensive layers. One approach is to build risk assessment frameworks that leverage a model's internal states to provide an early warning if it is about to generate insecure code~\cite{huang2025risk}. Another is to use program analysis to detect potential vulnerabilities; for instance, dataflow analysis can trace the flow of tainted data~\cite{wang2024llmdfa}, and graph-based models can detect defects at the repository level~\cite{tao2025code}. Proactively, models can be explicitly taught secure coding practices through specialized fine-tuning with feedback from static analysis tools, or prompted to perform explicit security reasoning before generating code to reduce the likelihood of introducing common vulnerabilities~\cite{saqib2025teaching, yao2025training, liu2025purpcode}.

The most stringent level of software quality is addressed by \textit{Formal Verification \& Safety Assurance}. Here, AI is being developed to generate code that is provably correct, a critical need in safety-critical domains. Systems like Alphaverus can bootstrap the generation of formally verified code by using formal proofs as an intermediate representation to guide code synthesis, even in domains where training data is scarce~\cite{aggarwal2024alphaverus, zhang2025building}. This verification-aware approach is also being explored in automated program repair to increase the reliability of generated patches~\cite{li2025patchpilot}. For dynamic systems such as autonomous driving software, a simulation-guided approach is used, where AI-generated code is rigorously tested within a high-fidelity simulator so that its safety and reliability could be ensured before deployment~\cite{nouri2025simulation}.

\section{Challenges and Future Directions}
\label{sec:challenges_and_future}

While the integration of LLMs has catalyzed significant advances in software engineering, the path toward fully autonomous and reliable agentic systems is fraught with fundamental challenges.
These obstacles, spanning scalability, evaluation, and responsible deployment, are not merely incremental hurdles but rather define the critical research frontiers for the field.
This section outlines these principal challenges and, in turn, proposes corresponding future research directions poised to overcome them, shaping the next generation of software engineering AI.

\subsection{Scalability: From Token Limits to Hierarchical Cognition}
A primary challenge is one of scale, where the architectural constraints of current LLMs prevent meaningful engagement with real-world software complexity.
While context windows have expanded, they remain insufficient for enterprise systems where codebases routinely exceed millions of lines of code.
This limitation manifests as project amnesia, where agents lose track of high-level architectural patterns and design conventions during extended tasks.
Current mitigation strategies, such as vector database retrieval, provide only partial solutions, as they fetch isolated snippets of context without comprehending the intricate relationships that define a software architecture.
At this scale, the computational economics of iterative refinement also become prohibitive, hindering the application of agents to complex, system-wide tasks.

To overcome these scalability limitations, future research must pivot from expanding linear context windows to developing hierarchical cognitive architectures for code understanding.
This necessitates a paradigm shift from processing code as a flat sequence of tokens to reasoning over structured, multi-modal representations like Code Property Graphs (CPGs).
Such architectures must be coupled with advanced tiered memory systems, such as combining a volatile short-term memory for immediate context with a persistent long-term memory storing architectural knowledge and project conventions.
By learning to navigate these compressed, structured representations, agents can transcend token limits to reason about system-wide properties, marking a crucial transition from agents that merely write code to agents that truly comprehend software.

\subsection{Evaluation: From Functional Correctness to Production Readiness}
The field currently faces an evaluation crisis, where prevailing benchmarks and metrics incentivize superficial correctness over the qualities that define production-ready code.
An over-reliance on pass@k on synthetic, function-level datasets like HumanEval creates a significant reality gap, failing to measure critical non-functional requirements such as security, performance, and maintainability.
Consequently, agents are optimized to hack the benchmark rather than to produce robust, high-quality software.
Recent analyses reveal that a majority of algorithmically successful solutions on benchmarks like SWE-Bench would fail in production due to the introduction of new security flaws, performance regressions, or violations of coding standards.
Without evaluation frameworks that mirror real software lifecycle concerns, progress remains illusory.

Addressing this evaluation gap requires a move toward holistic, lifecycle-aware benchmarking.
The next generation of evaluation frameworks should be built around dynamic, digital twin environments that simulate the entire software development lifecycle.
Within these sandboxed environments, agents must be assessed on a suite of production-oriented criteria, including the introduction of security vulnerabilities, performance regressions, and the accumulation of technical debt.
Furthermore, evaluation must incorporate economic and operational metrics, such as the computational cost of a solution and the human effort required for review.
Developing such comprehensive benchmarks will close the gap between academic research and industrial practice, ensuring that progress is measured by the delivery of truly production-ready software.

\subsection{Deployment: From Ethical Quagmires to Responsible Collaboration}
The deployment of autonomous agents introduces a host of ethical, legal, and resource-related challenges that create an accountability vacuum and threaten sustainable adoption.
Intellectual property frameworks are unprepared to resolve attribution for code co-created by humans, open-source libraries, and AI, creating significant legal uncertainty.
Operationally, the exponential scaling of computational costs presents an unsustainable model, while the workforce faces potential deskilling.
Most critically, when AI-generated code fails, the question of liability is diffused among the developer who prompted the agent, the organization that deployed it, and the creators of the model, creating a trilemma that obstructs adoption in mission-critical industries.

Navigating these challenges requires a concerted research effort toward responsible, sustainable, and collaborative AI systems.
To establish accountability, agents must be built with inherent mechanisms for transparency, such as the ability to generate verifiable audit trails that explain design rationale and trace code to its sources.
To ensure sustainability, research must prioritize resource-frugal agent architectures and cost-aware planning algorithms.
Finally, the paradigm must shift from pure automation to human-centric augmentation, framing AI as a scaffolding tool that empowers and upskills developers.
By designing agents as accountable, sustainable, and collaborative partners, we can build the foundation of trust required for their ethical and widespread adoption in the software engineering ecosystem.

\section{Challenges and Future Directions}
\label{sec:challenges_and_future}

While the integration of LLMs has catalyzed significant advances in software engineering, the path toward fully autonomous and reliable agentic systems is fraught with fundamental challenges.
These obstacles are not merely incremental hurdles but rather define the critical research frontiers for the field.
This section outlines these principal challenges and proposes corresponding future research directions poised to overcome them, shaping the next generation of software engineering AI.

\subsection{Scalability: From Token Limits to Hierarchical Cognition}
A primary challenge is one of scale, where the architectural constraints of current LLMs prevent meaningful engagement with real-world software complexity.
While context windows have expanded to millions of tokens, they remain insufficient for enterprise systems where codebases routinely exceed tens of millions of lines of code distributed across thousands of modules.
This limitation manifests as \textit{project amnesia}, where agents lose track of high-level architectural patterns, design conventions, and cross-module dependencies during extended tasks.
Current mitigation strategies, such as vector database retrieval and code summarization, provide only partial solutions, as they fetch isolated snippets of context without comprehending the intricate relationships that define a software architecture.
At this scale, the computational economics of iterative refinement also become prohibitive, hindering the application of agents to complex, system-wide tasks that may require dozens of edit-compile-test cycles.

To overcome these scalability limitations, future research must pivot from expanding linear context windows to developing \textit{hierarchical cognitive architectures} for code understanding.
This necessitates a paradigm shift from processing code as a flat sequence of tokens to reasoning over structured, multi-modal representations such as Abstract Syntax Trees (ASTs), Control Flow Graphs (CFGs), and Code Property Graphs (CPGs).
Such architectures must be coupled with advanced tiered memory systems, combining a volatile short-term memory for immediate context with a persistent long-term memory storing architectural knowledge, design patterns, and project conventions.
Promising directions include graph neural networks for repository-level code representation, hierarchical attention mechanisms that operate at multiple levels of abstraction, and neuro-symbolic approaches that integrate symbolic reasoning with neural code generation.
By learning to navigate these compressed, structured representations, agents can transcend token limits to reason about system-wide properties such as modularity, coupling, and architectural integrity, marking a crucial transition from agents that merely write code to agents that truly comprehend software systems.

\subsection{Evaluation: From Functional Correctness to Production Readiness}
The field currently faces an evaluation crisis, where prevailing benchmarks and metrics incentivize superficial correctness over the qualities that define production-ready code.
An over-reliance on pass@k metrics on synthetic, function-level datasets like HumanEval creates a significant \textit{reality gap}, failing to measure critical non-functional requirements such as security, performance, maintainability, and adherence to coding standards.
Consequently, agents are optimized to maximize benchmark performance rather than to produce robust, high-quality software suitable for deployment in production environments.
Recent analyses reveal that a majority of algorithmically correct solutions on benchmarks like SWE-Bench would fail code review in production due to the introduction of security vulnerabilities, performance regressions, violations of coding standards, or accumulation of technical debt.
Furthermore, existing benchmarks often use static test suites that can be gamed through memorization or superficial pattern matching, rather than measuring genuine problem-solving capability.
Without evaluation frameworks that mirror real software lifecycle concerns and incorporate adversarial testing, progress remains illusory.

Addressing this evaluation gap requires a move toward \textit{holistic, lifecycle-aware benchmarking} that captures the full spectrum of software quality attributes.
The next generation of evaluation frameworks should be built around dynamic, digital twin environments that simulate the entire software development lifecycle, from requirements elicitation through deployment and maintenance.
Within these sandboxed environments, agents must be assessed on a comprehensive suite of production-oriented criteria, including the absence of security vulnerabilities (e.g., CWE compliance), performance characteristics (e.g., time and space complexity), code quality metrics (e.g., maintainability index, cyclomatic complexity), and alignment with project-specific coding standards.
Furthermore, evaluation must incorporate economic and operational metrics, such as the computational cost of generating a solution, the human effort required for review and integration, and the long-term maintenance burden.
Promising directions include adversarial test generation to assess robustness, mutation testing to evaluate test suite adequacy, and longitudinal studies that track code evolution over time.
Developing such comprehensive benchmarks, potentially through partnerships with industry to access real production codebases and issue trackers, will close the gap between academic research and industrial practice, ensuring that progress is measured by the delivery of truly production-ready software.

\subsection{Domain Adaptation: From General Patterns to Specialized Expertise}
A critical yet underexplored challenge lies in the domain adaptation problem, where agents trained primarily on public repositories struggle to generalize to specialized domains with distinct programming paradigms, legacy systems, or proprietary frameworks.
While current LLMs demonstrate impressive performance on mainstream languages like Python and JavaScript, they exhibit significant degradation when confronted with domain-specific languages (DSLs), legacy codebases written in COBOL or Fortran, or highly specialized domains such as embedded systems, high-performance computing, or formal verification.
This limitation stems from the imbalanced distribution of training data, where popular languages and frameworks dominate the corpus while niche but critical domains remain severely underrepresented.
Furthermore, different software domains embody fundamentally different design philosophies: real-time systems prioritize deterministic timing over code elegance, safety-critical systems demand formal verification over rapid prototyping, and high-performance computing requires deep hardware awareness absent in web development.
Agents that fail to internalize these domain-specific constraints and idioms produce code that, while syntactically correct, violates fundamental domain principles and is unsuitable for deployment.

Overcoming the domain adaptation challenge requires developing \textit{domain-aware agent architectures} that can rapidly specialize to new programming contexts with minimal supervision.
Future research should explore few-shot and zero-shot domain adaptation techniques, enabling agents to quickly internalize domain-specific patterns from limited examples, documentation, or expert demonstrations.
This may involve meta-learning approaches that train agents to adapt quickly across diverse programming paradigms, transfer learning strategies that leverage knowledge from related domains, or retrieval-augmented generation systems that dynamically incorporate domain-specific knowledge from curated repositories.
For legacy system modernization, a particularly promising direction is the development of \textit{hybrid neuro-symbolic agents} that combine neural code generation with symbolic reasoning engines to enforce hard constraints derived from domain specifications.
Additionally, building high-quality, domain-specific datasets and benchmarks for underrepresented areas, such as embedded systems programming or scientific computing, will be essential to drive progress.
Ultimately, the goal is to develop agents that are not merely polyglot programmers but true domain specialists capable of understanding and respecting the unique constraints, idioms, and quality standards of each software engineering context.

\subsection{Multi-Agent Coordination: From Individual Capability to Collective Intelligence}
As software engineering tasks grow in complexity, spanning multiple subsystems, programming languages, and architectural layers, the limitations of single-agent systems become increasingly apparent.
While individual agents may excel at localized tasks such as function generation or bug fixing, they struggle with system-level challenges that require coordinated reasoning across multiple components, such as refactoring a distributed system, implementing a cross-cutting feature, or resolving conflicts between concurrent modifications.
The emerging paradigm of multi-agent systems, where specialized agents collaborate to tackle complex tasks, introduces a new class of coordination challenges.
These include establishing effective communication protocols between agents with different specializations, preventing redundant or conflicting modifications when multiple agents edit the same codebase, managing dependencies and task ordering when subtasks have complex precedence constraints, and allocating computational resources efficiently across the agent collective.
Current multi-agent frameworks often rely on simple coordination mechanisms such as sequential pipelines or centralized orchestration, which fail to capture the dynamic, iterative, and often non-linear nature of real software development workflows.

Advancing multi-agent software engineering requires developing \textit{sophisticated coordination mechanisms} that enable true collective intelligence.
Future research should explore decentralized coordination protocols inspired by distributed systems and multi-agent reinforcement learning, allowing agents to negotiate task allocation, resolve conflicts, and dynamically reorganize based on task requirements without centralized control.
This includes developing shared memory architectures and communication standards that enable agents to maintain consistent views of the codebase and project state, as well as conflict resolution algorithms that can automatically merge or arbitrate between competing code modifications.
A particularly promising direction is the integration of formal methods for coordination, such as using temporal logic to specify and verify coordination protocols or employing distributed consensus algorithms to ensure consistency.
Additionally, research should investigate role specialization strategies that go beyond simple task-based division to create agents with complementary cognitive capabilities, such as pairing a creative code generator with a rigorous verifier, or combining a bottom-up implementer with a top-down architect.
To evaluate these systems, new benchmarks are needed that measure not just task success but coordination efficiency, communication overhead, and the ability to gracefully handle dynamic task requirements.
By developing agents that can truly collaborate rather than merely coexist, we can tackle the scale and complexity of modern software systems that exceed the capabilities of any individual agent.

\subsection{Continuous Learning: From Static Models to Evolving Expertise}
A fundamental limitation of current LLM-based agents is their static nature: once deployed, they remain frozen in time, unable to learn from new technologies, evolving best practices, or project-specific feedback.
This creates a \textit{knowledge drift problem}, where agents trained on historical data gradually become obsolete as programming languages introduce new features, frameworks release breaking changes, and community best practices evolve.
For instance, an agent trained before the introduction of async/await syntax in JavaScript or the major architectural changes in React Hooks would produce outdated code patterns.
Moreover, agents cannot currently learn from their own mistakes or incorporate human feedback at scale, missing crucial opportunities for improvement through interaction.
This static learning paradigm is particularly problematic in rapidly evolving domains such as web development, where new frameworks and best practices emerge constantly, and in specialized enterprise contexts, where agents must adapt to proprietary codebases, internal libraries, and organization-specific coding standards.
Without continuous learning capabilities, agents require expensive and frequent retraining cycles, limiting their long-term utility and cost-effectiveness.

Addressing this challenge requires developing \textit{self-evolving agent architectures} with continuous learning capabilities that enable adaptation to new contexts without catastrophic forgetting.
Future research should explore online learning techniques that allow agents to incrementally update their knowledge from user corrections, code review feedback, and execution outcomes, while preserving core competencies.
This includes developing memory-efficient parameter update strategies such as adapter layers or low-rank adaptation (LoRA), which enable selective fine-tuning without full model retraining.
A particularly promising direction is \textit{reinforcement learning from human feedback} (RLHF) in the software engineering loop, where agents continuously improve by learning from developer interactions, code review outcomes, and post-deployment performance metrics.
Additionally, agents should maintain versioned knowledge bases that track the evolution of programming language features, library APIs, and best practices over time, enabling them to generate code appropriate for specific language versions or project configurations.
For organization-specific adaptation, research should investigate federated learning approaches that allow agents to learn from proprietary codebases while preserving intellectual property and privacy.
Ultimately, the goal is to develop agents that behave more like human developers who continuously learn throughout their careers, rather than static tools that become obsolete and require replacement.

\subsection{Responsible Deployment: From Ethical Quagmires to Trustworthy Collaboration}
The deployment of autonomous agents introduces a host of ethical, legal, and societal challenges that create an accountability vacuum and threaten sustainable adoption.
Intellectual property frameworks are unprepared to resolve attribution for code co-created by humans, open-source libraries, and AI, creating significant legal uncertainty that has already manifested in high-profile lawsuits.
When AI-generated code inadvertently incorporates copyrighted snippets or violates software licenses, determining liability becomes a complex trilemma involving the developer who prompted the agent, the organization that deployed it, and the creators of the underlying model.
Operationally, the exponential scaling of computational costs for advanced agents presents an unsustainable model that risks creating a two-tiered ecosystem where only well-resourced organizations can afford cutting-edge AI assistance.
Most critically, the workforce faces potential deskilling as developers increasingly rely on AI for tasks that were previously learning opportunities, raising concerns about the long-term health of the software engineering profession.
Additionally, the opacity of LLM decision-making creates trust barriers in safety-critical domains where code must be auditable and explainable.
These multifaceted challenges, if left unaddressed, risk undermining the transformative potential of AI in software engineering.

Navigating these challenges requires a concerted research effort toward \textit{responsible, sustainable, and trustworthy AI systems} that prioritize transparency, accountability, and human empowerment.
To establish accountability, agents must be built with inherent mechanisms for transparency, such as the ability to generate verifiable audit trails that explain design rationale, cite sources for generated code, and trace decisions to specific training data or retrieval sources.
This includes developing explainable AI techniques that can provide human-understandable justifications for code suggestions, as well as provenance tracking systems that document the lineage of generated code to facilitate license compliance and attribution.
To ensure sustainability, research must prioritize resource-frugal agent architectures, such as mixture-of-experts models that selectively activate subsets of parameters, and cost-aware planning algorithms that balance solution quality against computational expense.
For workforce considerations, the paradigm must shift from pure automation to \textit{human-centric augmentation}, framing AI as scaffolding that empowers and upskills developers rather than replaces them.
This involves designing agents that provide educational feedback explaining their reasoning, deliberately expose users to underlying concepts rather than hiding complexity, and adapt their level of assistance based on developer expertise.
Furthermore, establishing industry standards and governance frameworks for AI-generated code, including mandatory disclosure requirements, liability frameworks, and ethical guidelines, will be essential for building societal trust.
By designing agents as accountable, sustainable, and collaborative partners that enhance rather than diminish human capability, we can build the foundation of trust required for their ethical and widespread adoption in the software engineering ecosystem.
\section{Conclusion}
\label{sec:conclu}
This survey has presented the first comprehensive analysis connecting benchmarks and solutions in LLM-empowered software engineering, addressing a critical gap in understanding how evaluation methodologies align with solution approaches. Through systematic analysis of 150+ papers, we have demonstrated the field's evolution from simple prompt engineering to sophisticated agentic systems incorporating planning, reasoning, memory, and tool augmentation. Our unified taxonomy and pipeline provide researchers with a roadmap for selecting optimal approaches across varying task complexities. The identification of 50+ benchmarks and their corresponding solution strategies establishes a foundation for standardized evaluation and comparison. Looking forward, the emergence of multi-agent collaboration, self-evolving systems, and formal verification integration promises to further revolutionize software engineering practices. This work serves as an essential resource for advancing LLM-based software engineering toward more autonomous, reliable, and intelligent systems that can tackle increasingly real-world challenges.


\bibliographystyle{IEEEtran}
\bibliography{IEEEabrv,ref}

\end{document}